\documentclass[11pt]{article}

\usepackage{fullpage,amsmath,amsthm}
\usepackage{amssymb,latexsym}

\usepackage{times}

\newcommand{\myomit}[1]{}

\newcommand{\ket}[1]{|#1\rangle}
\newcommand{\bra}[1]{\langle#1|}
\newcommand{\ketbra}[2]{|#1\rangle\langle#2|}

\newcommand{\Tr}{\mbox{\rm Tr}}

\DeclareMathOperator{\tr}{tr}

\DeclareMathOperator{\poly}{poly}

\newcommand{\Pe}{\textsc{P}}
\newcommand{\NP}{\textsc{NP}}

\newcommand{\IP}{\textsc{IP}}

\newcommand{\MIP}{\textsc{MIP}}
\newcommand{\PSPACE}{\textsc{PSPACE}}

\newtheorem{theorem}{Theorem}
\newtheorem*{thm}{Theorem}
\newtheorem*{conj}{Conjecture}

\newtheorem{lemma}[theorem]{Lemma}

\newtheorem{claim}[theorem]{Claim}
\newtheorem{fact}[theorem]{Fact}
\newtheorem{corollary}[theorem]{Corollary}
\newtheorem{definition}[theorem]{Definition}

\newcommand{\be}{\begin{eqnarray}}
\newcommand{\ee}{\end{eqnarray}}

\newcommand{\Id}{\ensuremath{\mathop{\rm Id\,}\nolimits}}

\newcommand{\eps}{\varepsilon}

\newcommand{\ignore}[1]{}

\newcommand{\q}[1]{\ensuremath{\mathbf q_{#1}}}

\newcommand{\an}[1]{\ensuremath{\mathbf a_{#1}}}
\newcommand{\anp}[1]{\ensuremath{\mathbf a'_{#1}}}

\newcommand{\wtr}{\tilde{W}_{\q{r}}^{\an{r}}}
\newcommand{\vt}[1]{\ensuremath{\tilde{V}_{\q{r}}^{\an{#1}}}}
\newcommand{\wk}[1]{W_{\q{#1}}^{\an{#1}}}
\newcommand{\wkk}{\wk{k}}

\begin{document}

\thispagestyle{empty}

\title{Entangled games are hard to approximate}
\author{
Julia Kempe\thanks{Work partly done while at LRI, Univ. de Paris-Sud, Orsay. Partially supported by the
European Commission under the Integrated Project Qubit Applications (QAP) funded by the IST directorate as
Contract Number 015848, by an Alon Fellowship of the Israeli Higher Council of Academic Research and by a
grant of the Israeli Science Foundation.}\\School of Computer Science
\\Tel Aviv University\\ Tel Aviv, Israel\\
 \and
 Hirotada Kobayashi\thanks{
Supported by
the Strategic Information and Communications R\&D Promotion Programme No. 031303020
of the Ministry of Internal Affairs and Communications of Japan.}\\
 Principles of Informatics Research Division\\
  National Institute of Informatics\\
  Tokyo,
  Japan\\
 \and
 Keiji Matsumoto$^\dagger$\\
 Principles of Informatics Research Division\\
  National Institute of Informatics\\
  Tokyo,
  Japan\\
 \and
 Ben Toner\thanks{Part of this work was completed at Caltech.  Supported by the
National Science Foundation under Grants PHY-0456720 and CCF-0524828, by EU project QAP, by NWO VICI
project 639-023-302, and by the Dutch BSIK/BRICKS project.}
\\
CWI, Amsterdam\\ The Netherlands
 \and
Thomas Vidick\thanks{Work partly done while at LRI, Univ. de Paris-Sud, Orsay, and at DI, \'Ecole Normale
Sup\'erieure, Paris, France.}\\
Computer Science Division\\ University of California, Berkeley\\
USA
 }
\date{}

 \maketitle
\thispagestyle{empty}

\begin{abstract}
  We establish the first hardness results for the problem of computing the value of
  one-round games played by a verifier and a team of provers who can
  share quantum entanglement.
  In particular, we show that it is \NP-hard to approximate within
  an inverse polynomial the value of a one-round game with (i) quantum
  verifier and two entangled provers or (ii) classical verifier and
  three entangled provers. Previously it was not even known if computing the value exactly is \NP-hard.
  We also describe a mathematical
  conjecture, which, if true, would imply hardness of approximation to
  within a constant.

  We start our proof by describing two ways to modify classical
  multi-prover games to make them resistant to entangled provers. We
  then show that a strategy for the modified game that uses
  entanglement can be ``rounded'' to one that does not.  The results
  then follow from classical inapproximability bounds. Our work
  implies that, unless $\Pe=\NP$, the values of entangled-prover games
  cannot be computed by semidefinite programs that are polynomial in
  the size of the verifier's system, a method that has been successful
  for more restricted quantum games.
\end{abstract}
\newpage
\setcounter{page}{1}

\section{Introduction}

Multi-prover games have played a tremendous role in theoretical computer science over the last two decades.
In this setting, several provers, who are not allowed to communicate with each other during the game, exchange messages with a verifier according to a
prescribed protocol and try to convince him to accept. The {\em value} of a game is the maximum probability
with which the provers can achieve this, averaged over all the verifier's questions and possibly over the
shared randomness of the provers. The Cook-Levin Theorem implies that it is $\NP$-complete to compute the
value of such a game, where  the input is an explicit description of the game, i.e., a set of possible
questions, possible answers, a distribution on questions and acceptance predicates for the verifier. A lot of
research effort went into determining how hard it is to {\em approximate} the value of such games,
culminating in the celebrated PCP Theorem \cite{ALMSS98,AS98}, which shows that the value of a two-prover
one-round game with a constant number of possible answers is $\NP$-hard to approximate to within some
constant. This result has had wide-ranging applications, most notably in the field of hardness of
approximation, where it is the basis of many optimal results.


When considering multi-prover games in the quantum world, the laws of quantum mechanics allow for a
fascinating new effect: namely, the provers can share an arbitrary {\em entangled} state, on which
they may perform any local measurements they like to help them answer the verifier's questions. The fact that
entanglement can cause non-classical correlations is a familiar idea in quantum physics, introduced in a
seminal 1964 paper by Bell \cite{Bell}. Most importantly, there is no physical way to prevent provers from sharing
entanglement or to limit how much they have. Compare this to the restriction that the provers cannot communicate during the game, which can be enforced physically by separating the provers in space so that there is no time for a message to travel from one to the other. It is thus a natural and important question to ask how shared entanglement between the provers
influences the value of the game, as entanglement can allow for new strategies of the provers. Notice that
entanglement can potentially either make it easier or harder to approximate the value of a game, and it is a
wide open question which of these two effects actually takes place. For example, no algorithm---of any
complexity at all---is known to approximate the value of an arbitrary entangled-prover game.
One of the  most important questions in
this field, which we answer in this paper, has been to determine if it is hard or easy to compute the value
of entangled-prover games.

Two recent results give evidence that entangled-prover games might actually be computationally much {\em
easier} than their classical counterparts. First, Cleve et al.~\cite{CleveHTW04} showed that in the case of a
particular class of two-prover one-round games, XOR-games, the value when provers are entangled can
be computed (to exponential precision) in polynomial time. In contrast, H\aa stad \cite{Has01} showed that
for these games {\em without} entanglement it is $\NP$-hard to approximate the value to within some constant.
To prove their result, Cleve et al.~show that the maximization problem of the two provers can be written as a
semidefinite program (SDP) of polynomial size. It is well known that there are polynomial time algorithms to
find the optimum of such SDPs up to exponential precision, and hence there is a polynomial time algorithm to
compute the value of this game. More precisely, Cleve et al.~show that there is an SDP relaxation for the
value of the game with the property that its solution can be translated back into a protocol of the provers.
This is possible using an inner-product preserving embedding of vectors into two-outcome observables due to
Tsirelson \cite{Tsirelson:85b}, which works in the particular case of XOR-games. It has been a major open
question whether this result generalizes beyond XOR-games.

In a second recent result giving evidence that entangled-prover games are easy, Kempe, Regev and Toner
\cite{KempeRegevToner} show that even for the class of {\em unique} games (which contains the class of
XOR-games), an SDP-relaxation of the game gives a good approximation to its value. Hence, for unique games
there is a polynomial time algorithm to {\em approximate} the value of the game to within a constant.

An SDP-relaxation is not specific to XOR-games or unique games and can be written for all entangled
two-prover games.\footnote{In particular it will also be a relaxation for the value of the classical game
(which is not tight in this case, unless $\Pe=\NP$).} If the SDP is tight (as in the case of XOR-games) or
close to tight (as in the case of unique games) there is a polynomial time algorithm to compute or
approximate the value of the game. It was speculated that perhaps SDPs can compute or at least approximate
well the value of an entangled game for more general games. The semidefinite programming approach has been
widely successful whenever quantum communication is involved: for example Kitaev and Watrous \cite{KitWat00}
have shown that SDPs can exactly compute the value of {\em single}-prover quantum games, Gutoski and Watrous
proved that the value of quantum refereed games is as hard to compute as the value of classical refereed
games again via semidefinite programming~\cite{gutoski:_towar}, and Kitaev showed that the cheating
probability for quantum coin-flipping protocols \cite{KitaevCoin} can be computed by SDPs. Moreover,
Navascues et al. \cite{NavascuesPA07} recently gave a hierarchy of SDP relaxations to approximate the value
of an entangled two-prover game; yet no bounds on the quality of approximation have been proved and these
SDPs are in general not of polynomial size.

The major open question is thus to determine if it is easy or hard to compute or even to approximate the
value of general entangled-prover games. In particular, would it be possible that the value of such games could be computed or approximated by an SDP?

\paragraph{Our results.} In this paper we resolve the open question above by showing for the first time that
it is $\NP$-hard to compute the value of entangled multi-prover games in the quantum world. We need to
distinguish between two types of entangled games: on one hand one can still restrict the (possibly entangled)
provers to classical communication; we call such games {\em classical entangled games}. On the other hand one
can also allow the provers to communicate {\em quantum} messages with a {\em quantum verifier}; we call these
games {\em quantum entangled games}. In both cases the hardness of computing the value of the game with
entangled provers was previously not known,\footnote{Kobayashi and Matsumoto \cite{KoMa03} showed that when
the communication and the verifier are quantum, but the provers do not share any entanglement, then the
resulting games behave like classical games without entanglement, i.e., it is \NP-hard to approximate their
value to within a constant.} and we show \NP-hardness in two cases: for two-prover one-round {\em quantum
entangled games} (in the first part of the paper) and for three-prover one-round {\em classical entangled
games} (in the second part). Then we proceed to show that even {\em approximating} the value of these two
types of games is \NP-hard, thus giving the first hardness of approximation results.\footnote{Obviously the
hardness of computation result is implied by the hardness of approximation result. We include it nonetheless
in Sec. \ref{sec:QMIPcompute} for the quantum entangled games to illustrate the main ideas.} Our main result
can be stated as follows:

\begin{theorem}\label{thm:main}
There exists a polynomial $p$ such that it is \NP-hard to decide, for an explicitly given
\begin{enumerate}
\item two prover one-round quantum entangled game $G$ or \item three prover one-round classical entangled
game $G$,
\end{enumerate}
whether its value is $1$ or  $1-1/p(|G|)$.\footnote{See Section~\ref{sec:not} for a precise definition of the
size $|G|$ of $G$.}
\end{theorem}

This  theorem implies that no polynomial-time algorithm can compute the value of an entangled game to within
polynomial precision. Given the importance of SDPs in results on entangled games, the following immediate
corollary is of interest:

\begin{corollary}
The success probability of classical entangled $3$-prover or quantum entangled $2$-prover games cannot be computed by SDPs of polynomial size, unless $\Pe=\NP$.
\end{corollary}
The results above leave open the case of {\em two}-prover one-round {\em classical} entangled games. In the
third part of this paper we give a hardness result for this type of game which is stated precisely in Section
\ref{sec:IP} in the setting of {\em succinct} games and interactive proofs; here we just give a brief
overview. This third result has a slightly different flavor: we scale up to games with exponential number of
questions and answers, but given succinctly (i.e. the game is given by a description of the circuit of the
verifier of size polynomial in $\log |Q|$, the length of the questions). For these games we show that to
approximate the value to within an inverse polynomial (in $\log |Q|$) is at least as hard as to approximate
to within a constant the value of classical {\em single}-prover {\em multi-round} games with polynomial
rounds. Note that this is a better approximation than in the first two results of our paper (where the
approximation was an inverse polynomial in $|Q|$), but our hardness in this case is weaker than in the
previous two results. In particular, combining this with an adapted version of Shamir's result \cite{Sha92}
that $\IP=\PSPACE$, our result implies \PSPACE $\subseteq \MIP^*(2,1)_{1,1-\poly^{-1}}$. Again, no such
result was previously known for these games.

All three results turn out to have something in common---in the analysis of all three of them we show that
by enforcing certain tests we obtain sets of projectors (which characterize the strategy of the provers)
which pairwise {\em ``almost commute"}. From this condition we need to derive a classical strategy for the
original classical game, and we do this in a similar fashion in all three cases.

\medskip

\noindent {\bf Proof ideas and new techniques.}

\smallskip

{\bf Reduction:} We prove our $\NP$-hardness results by a reduction from the hardness of approximation result
for classical (non-entangled) games, as implied by the PCP Theorem, which we state in the language of games:

\begin{thm}[PCP Theorem \cite{ALMSS98,AS98}]\label{thm:PCP}
There is a constant $s<1$ such that it is $\NP$-hard to decide, given a two-prover one-round game with a
constant number of answers, whether its value is $1$ or  $\leq s$.
\end{thm}


We start with an instance of such a classical two-prover one-round game and modify it to a two-prover
one-round quantum entangled game (or a three-prover classical entangled game, in the second part of this
paper) with the property that the value of the new entangled game is at least as big as the value of the
original game. In other words, if the value of the original game is $1$, the value of the new game is still
$1$. To show that it is \NP-hard to {\em compute} the value of the entangled game we need to show that if the
value of the original game is below $s$ then the value of the new entangled game is {\em smaller} than $1$.
In particular, it suffices to show that if the value of the new entangled game is $1$, then the value of the
original game is also $1$. To show this, we use a successful strategy of the entangled provers to construct a
strategy in the original game that achieves a large value (see {\em Rounding} below).

Because we only need to show this when the new value is {\em exactly} $1$ our task is fairly easy once we
have established how to modify the game. It requires substantially more work to prove the hardness of
approximation result. We perform the same reduction as in the exact case, but now we need to show that if the
value of the original game is at most $s$, then the value of the new entangled game is bounded away from $1$
by an inverse polynomial. Equivalently, we have to show that if the value of the new entangled game is above
$1-\varepsilon$ for some inverse polynomially small $\eps$, then the value of the original classical game is
{\em larger} than $s$.

{\bf Modify the game to ``immunize" against entanglement:} An essential novel technique in our paper is the
design of the new games used in our reduction. We  design the new games in a way that limits the cheating
power of entangled provers. To this end---and this is a crucial difference to previous attempts to upper
bound the value of entangled games---we add an extra test to the game. This new test, which can be added
generically to {\em any} two-prover one-round game, significantly limits the use of entanglement by the
provers beyond its quality as shared randomness. We hope that this technique of ``immunizing" a game against
entanglement can be extracted to serve a wider purpose in other contexts where we want to limit the power of
entanglement, possibly with cryptographic applications.

In hindsight the fact that we need to modify the games comes as no surprise. Several classical games have
been analyzed in the past to show that without modification of the game, entanglement drastically increases
their value. One striking example is given by the Magic Square game \cite{Arvind:02}: Two classical players
can win this game with probability at most $17/18$. However, when given entanglement, the players can win
{\em perfectly}, i.e., they have a strategy that wins with probability $1$.

Our next novel element is the actual design of the new test. The difficulty is to show that entanglement does
not help the provers to coordinate their replies to increase the success probability. In the case of quantum
games (in the first part of this paper) our idea is to astutely use {\em quantum} messages and {\em quantum}
tests, and in particular a version of the SWAP-test, to enforce (approximately) that the provers do not
entangle the message register with the entangled state they share. This allows us to get conditions that
involve the provers' operators (describing their strategies) on two {\em different} questions. For this it is
crucial that the messages are quantum; we do not see any way to achieve this result for classical messages.

When we analyze classical entangled games (in the second part of our paper) we design a different test: we
modify the game by introducing a {\em third} player. We use the extra player to introduce a consistency test
that forces two of the provers to give the {\em same} answer. As a result, to pass this test, the two
original players can only use an entangled state of a specific form; it must be (approximately) {\em
extendable}, i.e., it must be the density matrix of a symmetric tripartite state. There are prior results
pointing to the potential usefulness of a third player to limit the cheating power of entanglement. For
example, two entangled provers can cheat in the Odd Cycle game of Ref.~\cite{CleveHTW04}, but if we add a
third prover, then entangled provers can perform no better than classical ones~\cite{toner:_monog}. Moreover,
after the completion of this work we have learned from A. Yao \cite{Yao:QIP07} about a way to add a third
player to the Magic Square game such that as a result the winning probability of entangled provers is
$\approx 0.94$.

For our third result on two-prover classical entangled games, our reduction has the same spirit and similar
analysis as in the previous two cases: here we start with a {\em single}-prover multi-round game and modify
it to a one-round game by introducing a second prover to prevent the first prover to entangle the answers of
subsequent rounds. Our modification here mimics a construction of \cite{CaiConLip94JCSS} used to prove that
$\PSPACE$ has (non-entangled) two-prover one-round systems.\footnote{In fact, we show that the
\cite{CaiConLip94JCSS} construction still remains sound even with entangled provers, albeit with a weaker
soundness than in the classical case.}

{\bf Rounding:} The extra quantum test (resp., the extra player) allows us to extract a mathematical
condition on the operations of the entangled players. More precisely it turns out that the projectors
corresponding to the various questions of the verifier pairwise ``almost commute'' in some sense or ``almost
do not disturb'' the entangled state. This means that the provers' actions are ``almost classical", in the
sense that they allow us to take any strategy in the entangled game and convert it back to a strategy in the
original classical game. We call this conversion {\em rounding} from a quantum solution to a classical
solution, in analogy to the rounding schemes used to convert a solution to an SDP relaxation to a solution of
the game. To explain the idea of our new rounding scheme, assume that the provers, when receiving a question
from the verifier, perform a projective measurement on their share of the entangled state depending on the
question, and answer with the outcome they get (it will turn out that this is essentially what the provers
can do, even when the game involves quantum communication). In the {\em exact} case, when the value of the
entangled quantum game is $1$, the measurements corresponding to different questions {\em commute} exactly.
Hence, there is a common basis in which the projectors corresponding to different answers are all diagonal
for all questions. In other words, for each question, the projectors simply define a partition of the basis
vectors. The probability that the provers give a certain pair of answers just corresponds to the size of the
overlap of the supports of the two corresponding projectors, i.e., to the number of basis vectors that are
contained in both of them. We can now construct a classical strategy for the original game, where the provers
use shared randomness to sample a basis vector, check which projector/partition contains it, and output the
corresponding answer. This classical strategy achieves exactly the same probability distribution on the
answers, and hence the same value of the game.

Matters complicate in the case where the value of the entangled game is $1-\eps$. Now, the provers'
measurements corresponding to different questions ``almost commute". To exploit this property in a rounding
scheme, imagine the following pre-processing step to eliminate entanglement from the strategy: Before the
game starts, the provers apply in sequence all possible measurements, corresponding to all possible
questions, on a share of the entangled state, and write down a list of all the answers they
obtain.\footnote{Obviously, the provers do not really need any entanglement to do this: all they have to do
is sample from the joint distribution that corresponds to the distribution of all the answers in this
sequence of measurements.} Then, during the game, when they receive a question from the verifier, they
respond with the corresponding answer in their list. Because the measurements almost commute, the answer to
any one particular question in this sequential measurement scheme are similarly distributed to the scenario
in the entangled game, where the prover only performs the one measurement corresponding to that question.
This can be seen by ``commuting" the corresponding projectors through the list of projectors in the
measurement, where each time we commute two operators we loose an $\eps$ in precision. As a result, also the
success probability of this new unentangled strategy is similar to the one in the entangled game, or at least
not too low.

{\bf A new mathematical challenge:} As mentioned above, our tests enforce an almost-commuting condition on
the operators of the provers. If they would commute exactly, they would be diagonal in a common basis, which
means that the strategy is essentially classical and does not use entanglement. If one could conclude that
the operators are {\em nearly diagonal} in some basis, one could again extract a classical strategy as in the
exact case. Hence we reduce proving {\em constant} hardness of approximation to the question whether one can
approximate our operators by commuting ones. This touches upon a deep question in operator algebra: {\em Do
almost commuting matrices nearly commute?} Here {\em almost commuting} means that the commutator is small in
some norm, and nearly commuting means that the matrices can be approximated by matrices that are diagonal in
some common basis. This famous question was asked for {\em two Hermitian} matrices by Halmos back in 1976
\cite{Halmos:conjecture}.\footnote{For the operator norm.} It was shown subsequently
\cite{Voiculescu:ex},\footnote{For a simpler, elegant proof see \cite{ExelLoring:89}.} using methods from
algebraic topology, that this conjecture is false for two {\em unitary} matrices. Then, Halmos' conjecture
was disproved in the case of three Hermitian matrices. Finally Halmos' conjecture was proved
\cite{Lin:commuting} by a ``long tortuous argument" \cite{Szarek:survey} using von Neumann algebras, almost
$20$ years after the conjecture had been publicized. In our case we reduce proving hardness of approximation
of the value of an entangled game to the conjecture for a set of pairwise almost commuting {\em projectors},
where the norm is the Frobenius norm $\|A\|^2_2=\Tr(A^\dagger A)$ (see Sec. \ref{sec:QMIPcompute}):
\begin{conj}
Let $W_1,\ldots ,W_n$ be $d$-dimensional projectors such that for some $\eps\geq 0$ for all $i,j\in
\{1,\ldots ,n\}$ $\frac{1}{d}\|W_iW_j-W_jW_i\|^2_2\leq \eps$. Then there exists a $\delta\geq 0$, and
pairwise commuting projectors $\tilde W_1,\ldots \tilde W_n$ such that $\frac{1}{d}\|W_i-\tilde W_i
\|^2_2\leq \delta$ for all $i \in \{1,\ldots ,n\}$.
\end{conj}
Our proof shows that the conjecture with a constant $\delta$ implies hardness of approximation of the value
of entangled games to within a {\em constant}, i.e., best possible. For two, three or a constant number of
projectors the conjecture is easy to prove for a constant $\delta$. We do not know if it is true in general.

\paragraph{Related work.}
A subset of the authors has obtained weaker results on harness of approximation of the value of entangled
two-prover {\em quantum} games, posted to the arXiv earlier \cite{KempeVidick}; the present paper includes
and supersedes these results. Since this paper had been made public, our techniques have already been applied
by \cite{yao:tsirelson} to show similar results for {\em binary} three-player one-round classical entangled
games. \cite{yao:tsirelson} also give a new upper-bound for the value of these games; or, as often called in
this context, they gave a new tripartite Tsirelson-inequality. After the completion of this work Cleve,
Gavinsky and Jain \cite{Cleve:NPinMIPstar} use a connection to private information retrieval schemes to show
that succinctly given binary entangled classical games can not be approximated in polynomial time. Their
result does not apply for explicitly given games, as it is based on an exponential expansion of the message
length. It uses very different techniques, and is not comparable to ours.

{\bf Structure:} The structure of this paper is as follows: In Section \ref{sec:not} we introduce the
necessary definitions and notations we use. In Section \ref{sec:QMIP} we prove our results on the
\NP-hardness of quantum entangled two-prover games. To flash out the ideas, we first prove hardness of {\em
computing} the value of such games, before showing hardness of approximation. In Section \ref{sec:MIP} we
show \NP-hardness of approximation for the value of three-prover classical entangled games, and in Section
\ref{sec:IP} we give our hardness results for two-prover classical entangled games. We discuss our results
and open questions in Section \ref{sec:rest}.

\section{Preliminaries}\label{sec:not}

We assume basic knowledge of quantum computation~\cite{nielsen&chuang:qc}.

\paragraph{Games.} In this paper we study multi-prover games, or cooperative games with imperfect information
(henceforth {\em games}). We will only deal with one-round games played by $N$ cooperative provers against a
verifier. For an integer $K$, denote $\{1,\dots ,K\}$ by $[K]$.

\begin{definition}\label{def:games} Let ${Q}$ and ${A}$ be integers. A game $G = G(N,\pi, V)$ is given by a set
$\bar Q=\{q_{i_1 \ldots i_N}\}_{(i_1 \ldots i_N) \in [{Q}]}$ of questions and $\bar A=\{a_{i_1 \ldots
i_N}\}_{(i_1 \ldots i_N) \in [{A}]}$ of \emph{answers}, together with a distribution $\pi: [Q]^N \to [0, 1]$,
and a function $V:[ A]^N \times [ Q]^N \rightarrow \{0,1\}$.\footnote{We write $V(\cdot, \cdot)$ as $V(\cdot
|\cdot)$ to clarify the role of the inputs.}  The value of the game is\footnote{We use a supremum because the
optimal strategies might not be finite in the case of entangled provers.}
 \begin{equation}\label{eq:value} \omega (G) = \sup
_{W_1,\ldots,W_N} \sum_{i_1,\ldots , i_N \in [Q]^N} \pi(i_1,\ldots,i_N)\sum_{j_1,\ldots,j_N \in [\bar A]^N}
\Pr(a_{j_1 \cdots j_N})V(a_{j_1\cdots j_N} |i_1\cdots i_N),
 \end{equation}
 where the $W_i$ are the prover's strategies, and the
probability $\Pr(a_{j_1\cdots j_N})=\Pr(W_1(i_1,r)\cdots W_N(i_N,r)=a_{j_1\cdots j_N})$ is taken over the
randomness of the provers.
\end{definition}

The game $G$ is played as follows: The verifier samples $i_1, \ldots ,i_N$ from $[Q]^N$ according to $\pi$,
and prepares a  question $q_{i_1 \cdots i_N} \in \bar Q$. He sends the $k$-th part of the question to prover
$k$ for $1 \leq k \leq N$ and receives the answer $a_{j_1 \cdots j_N} \in \bar A$ from the provers. The
provers win the game if $V(a_{j_1 \cdots j_N}| i_1 \cdots i_N)$ = 1; otherwise the verifier wins. The {\em
value} of a game is the maximum winning probability of the provers. The provers can agree on a strategy
before the game starts, but are not permitted to communicate after receiving questions.

We distinguish three different kinds of games, based on the classical or quantum nature of the verifier, the
provers, and the question and answer sets. A game $G$ will be called
\begin{itemize}
\item \emph{classical} if the verifier, the prover, and the question and answer sets are classical. In this
case $q_{i_1 \cdots i_N}=(q_1,\ldots ,q_N)$ and $a_{i_1 \cdots i_N}=(a_1,\ldots ,a_N)$ are $N$-tuples, i.e.,
the verifier simply sends $q_k$ to the $k$-th prover and receives $a_k$ from him. We identify $\bar Q$ with
$[Q]^N$, $\bar A$ with $[A]^N$, $i_k$ with $q_k$, and $j_k$ with $a_k$  and often write $Q$ for $[Q]$ and $A$
for $[A]$. The strategies $W_i$ are simply functions $W_i:Q \times R \rightarrow A$ where $R$ is some
arbitrary domain (``shared randomness"). In fact we can assume the strategies to be {\em deterministic}:
there is always some $r \in R$ that maximizes the winning probability and we can fix it in advance.

\item \emph{classical entangled} if the verifier, and the question and answer sets are classical, but the
provers are quantum, and are allowed to share an a priori entangled state $\ket{\Psi}$ of arbitrary
dimension. This increases the set of possible strategies to quantum operations performed on the prover's
share of the entangled state. By standard purification techniques (see, e.g, \cite{CleveHTW04}) one can
assume that each prover performs a projective measurement ${\cal W}_q=\{W_q^a\}_{a \in A}$ with outcomes in
$A$ (i.e., $\sum_{a \in A} W^a_q=\Id$ and $(W_q^a)^\dagger=W_q^a=(W_q^a)^2$), where we adopt the same
notational identifications as for classical games. We will use a superscript $*$ to indicate entangled-prover
games. The value $\omega^* (G)$ of such a game is given by Eq. (\ref{eq:value}) where the probability
$\Pr(a_1\,ldots ,a_N)=\bra \Psi (W_1)^{a_1}_{q_1} \otimes \cdots \otimes (W_N)^{a_N}_{q_N}\ket \Psi$.

\item \emph{quantum entangled} if both the verifier and the provers are quantum, and they exchange quantum
messages. We usually denote such a game by $G_q$. In that case $q_{i_1 \cdots i_N} \in \bar Q$ is a joint
density matrix and the verifier sends its $k$-th part to the $k$-th prover for $1 \leq k \leq N$ using a
quantum channel, possibly keeping a part in his own private register. After receiving as answer an
$N$-register quantum state $a_{j_1 \cdots j_N} \in \bar A$, where the $k$-th prover sends the $k$-th
register, the verifier performs a quantum operation $V'$ (which might depend on the questions in $[Q]^N$) on
the answer and his private space, followed by a measurement $\{\Pi_{acc},\Pi_{rej}\}$ of his first qubit. By
purification we can assume that the $k$th prover performs a unitary transformation $U_k$ on the message
register and his part of the entangled state $\ket{\Psi}$ and then sends the message register back to the
verifier. The value of an entangled-prover quantum game, $\omega_q^*$, is given by Eq. (\ref{eq:value}) where
$$\sum_{j_1,\ldots,j_N}\Pr(a_{j_1\cdots j_N})V(a_{j_1\cdots j_N} |i_1\ldots i_N) = \Tr(\Pi_{acc} V'
(U_1\otimes\cdots\otimes U_N)( q_{i_1 \cdots  i_N}\otimes\ketbra{\Psi}{\Psi})).$$
\end{itemize}

\paragraph{Input size.}
A game is described by $Q,A,\pi$ and $V$, and hence our complexity parameter, the size of the input, is
polynomial in $Q$ and $A$.\footnote{Here we always assume that $N$ is a constant.} We will always assume that
the description of the distribution $\pi$ is of polynomial size in ${Q}$. In the case of quantum games we
also have to take into account the size of a description of the question $q_{i_1\ldots i_N}$, and the
verification procedure $V'$, and the dimension of the answer $a_{j_1 \ldots j_N}$: we always assume that the
dimensions of $q_{i_1\ldots i_N}$ and $a_{j_1 \ldots j_N}$ are polynomial in ${Q}$ and ${A}$ and hence there
is a (classical) description of $q_{i_1\ldots i_N}$ and of $V'$ (which can be assumed to be a unitary of
polynomial dimension) of polynomial size in ${Q},{A}$.\footnote{In fact all games we consider also have a
circuit of size $\poly \log {Q}$ to prepare $q_{i_1 \ldots i_N}$ from $i_1,\ldots ,i_N$.}

\paragraph{Symmetric games.} For convenience we will work with symmetric distributions $\pi$. The next lemma
shows why this poses no restriction (we only need the case of $2$ provers).

\begin{lemma}\label{lem:symmetry1}
For every game $G=G(2,\pi,V)$ there is a game $G'=G(2,\pi',V')$ of the same value and twice as many
questions, such that $\pi'$ and $V'$ are symmetric under permutation of variables. Moreover there is an
optimal symmetric strategy for $G'$.
\end{lemma}

\begin{proof}
The verifier $V'$ in game $G'$ samples $q,q'$ from $\pi$. He adds an extra bit register to the questions and
with probability $1/2$ he sends $(q,1)$ to prover $1$ and $(q',2)$ to prover $2$, otherwise he swaps the two
questions. In the second case he swaps the received answers and in both cases applies the predicate $V$. For
the lower bound observe that if $S_1,S_2$ is a strategy for $G$, then the strategy for $G'$ where each prover
applies $S_i$ if his second message bit is $i$ fares as well as $S_1,S_2$ (and is symmetric). For the upper
bound note that from any strategy $S_A,S_B$ for $G'$ we can construct a strategy for $G$ that fares at least
as well, by choosing the better of either $S_A(\cdot,1),S_B(\cdot,2)$ or $S_B(\cdot,1),S_A(\cdot,2)$.
Moreover, $V'$ is obviously symmetric under permutation of question-answer pairs.
\end{proof}

\noindent In the case where the provers are allowed to share entanglement, we can assume that if $\pi$ and
$V$ have some symmetry, it is mirrored in the optimal prover's strategies:

\begin{lemma}\label{lem:symmetry}
Let $G=G(N,\pi,V)$ be a (classical or quantum) entangled-prover game, such that $\pi(i_1,\ldots,i_N)$ is
symmetric in $i_1,\ldots ,i_k$ and $V$ is symmetric under simultaneous permutation of the registers $1 \ldots
k$ of the questions $q_{i_1 \cdots i_N}$ and of the answers $a_{i_1 \cdots i_N}$ for $k \leq N$. Then given
any strategy $P_1,\ldots,P_N$ with entangled state $\ket{\Psi}$ that wins with probability $p$, there exists
a strategy $P'_1,\ldots, P'_N$ with entangled state $\ket{\Psi'}$ and winning probability $p$ such that
$P'_{1}=\ldots = P'_{k}$ and $\ket{\Psi'}$ is symmetric with respect to the provers $1,\ldots,k$.
\end{lemma}

\begin{proof}
Let $\mathfrak{S}_{k}$ be the set of permutations of $\{1,\ldots,k\}$ and assume, by appropriately padding
with extra qubits, that the first $k$ registers of $\ket{\Psi}$ have the same dimension. Define strategies
$P'_1,\ldots,P'_N$ as follows: the provers share the entangled state
$\ket{\Psi'}=\sum_{\sigma\in\mathfrak{S}_{k}}
\ket{\sigma(1)}\ldots\ket{\sigma(k)}\otimes\ket{\Psi^{\sigma}}$, where the register containing
$\ket{\sigma(i)}$ is given to prover $i$ and $\ket{\Psi^{\sigma}}$ is obtained from $\ket{\Psi}$ by swapping
the first $k$ registers according to $\sigma$. For $i \leq k$ prover $i$ measures the register containing
$\ket{\sigma(i)}$ and applies $P_{\sigma(i)}$. For $i>k$, $P'_i=P_i$. By symmetry of $\pi$ and $V$ this new
strategy achieves the same winning probability $p$, and $\ket{\Psi'}$ has the required symmetry properties.
\end{proof}

\section{Hardness of two-prover entangled quantum games}\label{sec:QMIP}

In this section we prove Theorem~\ref{thm:main} for the case of two-prover quantum entangled games. To better
quantify the dependence on the input size, we restate it as a separate result:

\begin{theorem}\label{thm:QMIPmain}
There is a constant $s_q>0$ such that it is \NP-hard to decide, given an two-prover quantum entangled  game,
whether its value is $1$ or less than $1-\eps$ for $\eps=\frac{s_q}{|Q|^4}$.
\end{theorem}

As mentioned in the introduction, we will prove this by a reduction from the PCP Theorem. However, to more
clearly and cleanly expose the ideas in this proof, we will first prove the simpler statement about
\NP-hardness of {\em computing} the value.

\subsection{\NP-hardness of computing the value of entangled quantum games}\label{sec:QMIPcompute}

\begin{theorem}\label{thm:QMIPcompute}
It is \NP-hard to decide, given an two-prover quantum  entangled game, whether its value is $1$.
\end{theorem}

We first describe how to modify a two-prover classical game $G_c(2,\pi,V)$ with questions $Q$ and answers $A$
to a two-prover {\em quantum} game of equal or higher value. We assume that the distribution $\pi(q,q')$ is
symmetric (as per Lemma \ref{lem:symmetry1}, at the expense of doubling the number of questions) and also
that there is a non-zero probability for each question  to be asked (otherwise we remove it from $Q$ without
affecting the value of the game).

\paragraph{The modified quantum game.}
In the constructed quantum game $G_q$ the verifier sends  quantum registers $\ket{q,0}_A$ and $\ket{q',0}_B$
to provers $A$ and $B$. We call the first part of this register the {\em question register} and the second
part the {\em answer register}. The answer register is initially in some designated state $\ket{0}$ and the
provers are expected to write the answers $a \in A$ to the question $q \in Q$ into this register and then
send both registers back. The verifier performs one of two tests, with equal probability:

\smallskip

\noindent{\bf Classical Test:} The verifier samples $(q,q')$ according to the distribution $\pi(q,q')$, and
sends $\ket{q,0}$ to prover $A$ and $\ket{q',0}$ to prover $B$. Upon receiving these registers from the
provers, he measures them and accepts if the results of the measurement of the question registers is $q,q'$
and the results of the measurement of the answer registers $a,a'$ would win the game $G_c$.

\smallskip

\noindent{\bf Quantum Test:} The verifier samples $(q,q')$ according to the distribution $\pi(q)\pi(q')$,
where $\pi(q)$ is the marginal of $\pi(q,q')$ and prepares the state
 \begin{equation}\label{eq:swapstate}\frac{1}{\sqrt{2}}
\left(\ket{0}\ket{q,0}_A\ket{q',0}_B+\ket{1}\ket{q',0}_A\ket{q,0}_B\right).
\end{equation}
He keeps the first qubit and sends question and answer registers to provers $A$ and $B$. Upon receiving these
registers from the provers, he performs a controlled-SWAP on registers $A$ and $B$ conditioned on the first
qubit being $\ket{1}$ (he swaps both the question and the answer register). Then he measures his qubit in the
basis $\{\ket{+},\ket{-}\}$\footnote{Or, equivalently, he performs a Hadamard transform and measures his
qubit in the standard basis.} and the question registers.  He accepts iff the results of the measurement of
the question registers is $q,q'$ and the outcome of the measurement of the first qubit is ``$+$".

{\em Remarks:} Note that the value $\omega_q^*(G_q)$ of the constructed game $G_q$ is obviously at least the
value of $G_c$: If the entangled quantum provers, controlled on the question, simply write the answer that
the classical unentangled provers would have given into the answer register, they always pass the quantum
test, and hence $\omega_q^*(G_q) \geq \omega(G_c)/2+1/2 \geq \omega(G_c)$.

Moreover the description of the quantum game has essentially the same size as the description of the
classical game, i.e. the complexity parameter is the same in both cases. The dimension of question and answer
registers is $|Q|$ and $|A|$ and the SWAP test only requires extra space that is no more than linear in the
number of qubits swapped.

Note that it is only the SWAP-test that is genuinely quantum, and allows us to show that the provers cannot
entangle too much the questions they receive  with the entangled state they share, by relating their actions
on two different messages. This test has been used in various settings in the past. In its most simple form
it was used in \cite{bcww:fp} to give a protocol for quantum fingerprinting. However, the test that we
perform here is a little more sophisticated, since it implements only a {\em partial} SWAP on the two message
registers, which might be entangled with the prover's private spaces and entanglement,  on which the verifier
in unable to perform the swapping. This partial swap has been used in \cite{KitWat00} to show parallelization
for QIP, and in \cite{KMY03} to prove the inclusion QMA$(3)\subset$QMA$(2)$, where the $2$ and $3$ refer to
the number of Merlins.

A last remark concerns the two different probability distributions used in the two tests. We really need to
change the distribution in the quantum test, because it gives us a commutation condition for {\em all}
operators of the provers, corresponding to all different questions. Otherwise, we would only obtain it for
pairs of questions $q,q'$ corresponding to a non-zero $\pi(q,q')$, which is not sufficient to round to a
classical strategy.

\paragraph{Existence of a good classical strategy.}

We now show that if the value of the quantum game is $1$, then there is a strategy for the classical game
that wins with probability $1$.
\begin{lemma}\label{lemma:compute}
If $\omega_q^*(G_q)=1$ then $\omega(G_c)=1$.
\end{lemma}
This implies that if the value of the classical game was less than $1$, then the value of the quantum game is
less than $1$. Since it is \NP-hard to distinguish whether the value of the classical game is $1$ or not, it
follows that it is \NP-hard to decide whether the value of the quantum game is $1$.

\begin{proof}[Proof of Lemma \ref{lemma:compute}:]
Consider a maximizing strategy, which in particular passes the quantum test with certainty.\footnote{Strictly
speaking it could be that such a strategy exists only in the limit of infinite entanglement, so we would have
to use a strategy that achieves success probability arbitrarily close to $1$. Since in this part we only give
the ideas of the rigorous proof in Section \ref{sec:QMIPapprox}, we ignore this issue.} Note that if it were
not for the controlled-SWAP the game would be essentially an entangled {\em classical} game, because question
and answer registers are prepared in a classical state and are immediately measured when received by the
verifier. We first show that the strategy of the provers is indeed essentially a classical entangled
strategy.

\begin{claim}\label{claim:classical}
There is a shared bipartite state $\ket{\Psi}_{AB}$ and for each question $q \in Q$ a set of projectors
$\{W_q^a\}_{a \in A}$ acting on each prover's half of $\ket{\Psi}$ with $\sum_{a \in A} W_q^a=\Id$ such that
each provers' transformation can be written as $\ket{q}\ket{0}\ket{\Psi} \rightarrow \ket{q}\sum_a\ket{a}
W_q^a\ket{\Psi}$ and the probability that the verifier measures $a,a'$ in the answer registers given he
sampled $q,q'$ in the classical test is
$$p_q(a,a'|q,q')=\|W_q^a \otimes W_{q'}^{a'} \ket{\Psi}_{AB}\|^2.$$
\end{claim}

\begin{proof}
At the beginning of the protocol the provers share some  entangled state $\ket{\Psi'}$ (including their
private workspace). From Lemma \ref{lem:symmetry} we can assume that the strategies  in the quantum game are
symmetric, i.e., that $A$ and $B$ apply the same unitary transformation $U$. Since the provers pass the
quantum test perfectly it means that they do not change the question register. Hence it is easy to see that
$U$ is block-diagonal and can be written as $U=\sum_{q}\ket{q}\bra{q}\otimes U_{q}$ where $U_{q}$ acts on the
answer register and half of $\ket{\Psi'}$. Define the operators $\tilde{W}_q^a=\bra{a}U_q\ket{0}$, where
$\ket{0}$ and $\ket{a}$ only act on the answer register, not on $\ket{\Psi'}$, i.e.
$U_q\ket{0}\ket{\Psi'}=\sum_a \ket{a} \tilde W_q^a \ket{\Psi'}$. Then it follows that $\sum_a (\tilde
W_q^a)^\dagger \tilde W_q^a=\Id$, meaning that $\tilde W_q^a$ are superoperators acting on a part of
$\ket{\Psi'}$. By standard arguments we can now enlarge the system to a state $\ket{\Psi}$ such that we can
replace the $\tilde W_q^a$ by projectors $W_q^a$ which give exactly the same outcome probabilities.
\end{proof}

We now derive the crucial condition that allows us to define a good classical strategy.

\begin{claim}\label{lemma:commute1}
$$\forall q,q',a,a' \quad W_q^a \otimes W_{q'}^{a'}\ket{\Psi}=W_{q'}^{a'}\otimes W_q^a \ket{\Psi}.$$
\end{claim}

\begin{proof}
After the controlled-SWAP and the measurement of the question registers as $q,q'$, the remaining state of the
entire system can be described as
\begin{align*}&\frac{1}{\sqrt{2}}\sum_{a,a'}\ket{a}\ket{a'}\left(\ket{0}(W_q^a \otimes W_{q'}^{a'})
\ket{\Psi}+\ket{1}(W_{q'}^{a'}\otimes W_q^a)
\ket{\Psi}\right)\\&=\frac{1}{2}\sum_{a,a'}\ket{a}\ket{a'}\left(\ket{+}(W_q^a \otimes W_{q'}^{a'}
+W_{q'}^{a'}\otimes W_q^a) \ket{\Psi} +\ket{-}(W_q^a \otimes W_{q'}^{a'} -W_{q'}^{a'}\otimes W_q^a)
\ket{\Psi}\right)
\end{align*}
and hence the probability to measure ``$-$" in the extra qubit is $\frac{1}{4}\sum_{a,a'}\|(W_q^a \otimes
W_{q'}^{a'} -W_{q'}^{a'}\otimes W_q^a) \ket{\Psi}\|^2$ which must be $0$ since the provers pass the quantum
test with certainty.
\end{proof}
\noindent{\bf Rounding:} This property of the projectors can be expressed in a different fashion. Assume for
simplicity that the shared state is maximally entangled, i.e.,
$\ket{\Psi}=\frac{1}{\sqrt{d}}\sum_{i=1}^d\ket{i}_A\ket{i}_B$, and that all projectors are real. Then for any
such projectors $W,W'$ we have $\|W \otimes W' \ket{\Psi}\|^2=\frac{1}{d} \|WW'\|_F^2$, where
$\|A\|^2_F=\Tr(A^\dagger A)$ is the Frobenius norm. The condition in Claim \ref{lemma:commute1} can be
rewritten as $\frac{1}{d}\|W_q^a W_{q'}^{a'}-W_{q'}^{a'} W_q^a\|_F=0$, i.e. the two projectors {\em commute}.
Hence, in some basis $\{\ket{e_i}\}_{i=1}^d$, all $W_q^a$ are diagonal matrices with only $1$ and $0$ on the
diagonal. In other words, each projector simply defines a {\em partition} of the basis vectors, and
$p(aa'|qq')=\frac{1}{d}\|W_q^aW_{q'}^{a'}\|^2_F$ just measures the relative {\em overlap} of the two
partitions. With this in mind we can easily design a classical randomized strategy for $G_c$ with the same
success probability. The provers sample a shared random number $i \in \{1,\ldots,d\}$. When receiving
question $q$  they answer with $a$  such that the basis vector $\ket{e_i}$ is in the support of $W_q^a$.

This proof can be generalized to an arbitrary shared state $\ket{\Psi}$ and general projectors. We will not
give the full details  (in any case Thm. \ref{thm:QMIPcompute} follows from Thm. \ref{thm:QMIPmain}), but the
way to prove this is to define a diagonal real positive matrix $D$ with the Schmidt-coefficients of
$\ket{\Psi}$ in the diagonal. Then $\|W \otimes W' \ket{\Psi}\|^2= \|WDW'^T\|_F^2$, where the elements on the
diagonal of $D$ can be thought of as weights, and the condition in Claim \ref{lemma:commute1} becomes
$\|W_q^a D (W_{q'}^{a'})^T-W_{q'}^{a'}D (W_q^a)^T\|_F=0$. Moreover, following the same ideas as used in Claim
\ref{claim:commute} to show Eq. (\ref{eq:commute2}), we obtain $\|W_q^a D -D (W_q^a)^T\|_F=0$. Together these
conditions imply $W_q^a W_{q'}^{a'}D=W_{q'}^{a'}W_q^aD$, i.e. the two projectors commute over the space where
$D$ is non-zero. The classical strategy is now a weighted version of the strategy outlined in the case of a
maximally entangled shared state.
\end{proof}

\subsection{\NP-hardness of approximating the value of entangled quantum games}\label{sec:QMIPapprox}

With the intuitions obtained so far we can now tackle the harder case of hardness of approximation. First a
quick overview. We modify the game in exactly the same way as before. To prove Theorem \ref{thm:QMIPmain} we
now need to show, for $s$ from the PCP Theorem:

\begin{lemma}\label{lemma:mainQMIP}
If $\omega_q^*(G_q)>1-\eps$ then $\omega(G_c)> s$.
\end{lemma}
This implies that if the value of the classical game was less than $s$, then the value of the quantum game is
less than $ 1-\eps$. Since, from the PCP Theorem it is \NP-hard to distinguish whether the value of the
classical game is $1$ or less than $s$, it follows that it is \NP-hard to decide whether the value of the
entangled quantum game is $1$ or below $1-\eps$.

To prove Lemma \ref{lemma:mainQMIP}, we first show that the strategies of the provers are essentially
projective measurements (Claim \ref{claim:classicalQMIP}). We then extract the ``almost commuting" conditions
on the operators of the provers (Claim \ref{claim:commute}), which allow us to give a good strategy for the
original game.

\begin{proof}[Proof of Lemma \ref{lemma:mainQMIP}]
Consider a maximizing strategy.\footnote{Since it could be that the value of the game is only achieved in the
limit of infinite entanglement we in fact consider a strategy with finite entanglement that has success
probability $1-\eps-\delta$ for some arbitrarily small $\delta$. We will not write this $\delta$ in what
follows, but the proof goes through for small enough $\delta$, for instance $\delta=O(\eps)$.} It must pass
each of the two tests with probability at least $1-2\eps$. Again it is (approximately) true that the strategy
of the provers is essentially an entangled {\em classical} strategy.
\begin{claim}\label{claim:classicalQMIP}
There is a shared bipartite state $\ket{\Psi}_{AB}$ and for each question $q \in Q$ a set of projectors
$\{W_q^a\}_{a \in A}$ acting on each prover's half of $\ket{\Psi}$ with $\sum_{a \in A} W_q^a=\Id$  such that
if we replace each prover's transformation by $\ket{q}\ket{0}\ket{\Psi} \rightarrow \ket{q}\sum_a
\ket{a}W_q^a \ket{\Psi}$ then the probability to pass each of the tests is at least $1-6 \eps$ and the
probability distribution on the answers in the classical test is given by $$p_q(aa'|qq')=\|W_q^a \otimes
W_{q'}^{a'} \ket{\Psi}\|^2.$$
\end{claim}
\begin{proof}
As in the proof of Claim \ref{claim:classical} the provers apply the same unitary transformation $U$, which
now is not exactly block-diagonal, but in general can be written as $U=\sum_{q,\tilde q \in Q} \ket{\tilde
q}\bra{q}\otimes U_{q\tilde q}$. Because the verifier in both the classical and the quantum test measures
$q,q'$ in the answer register with probability at least $1-2\eps$, this implies that
$$ {\rm E}_{(q,q')}\left[\sum_{ \tilde q\neq q}\sum_{\tilde q'\neq q'}\| U_{q\tilde q}
\otimes U_{\tilde q'q'}\ket{0}_A\ket{0}_B\ket{\Psi'}_{AB} \|^2\right]\leq 2\eps,$$ for both when $(q,q')$ is
sampled according to  $\pi(q,q')$ (from the classical test) or according to $\pi(q)\pi(q')$ (from the quantum
test), where we have used symmetry of $\ket{\Psi'}$ for $\|\frac{1}{\sqrt{2}}(\ket{0} U_{q\tilde q} \otimes
U_{\tilde q'q'}+\ket{1} U_{\tilde q'q'} \otimes U_{q\tilde q}) \ket{0}_A\ket{0}_B\ket{\Psi'}_{AB}
\|^2=\|U_{q\tilde q} \otimes U_{\tilde q'q'}\ket{0}_A\ket{0}_B\ket{\Psi'}_{AB} \|^2$.

We approximate $U$ by a block-diagonal unitary operator $O_U$ as follows: extend each prover's private space
by registers $A'$ and $B'$ of dimension $|Q|+1$, initialized to $\ket{0}_{A'}$ and $\ket{0}_{B'}$ and let
$O_U= \sum_q \ketbra{q}{q} \otimes T_q$, where the unitary matrix $T_q$ acts on half of the entangled state
and the answer register (together shortened as $\ket{\cdot}$) and $A'$ as
$$T_q \ket{\cdot}
\ket{0}_{A'}=U_{qq}\ket{\cdot} \ket{0}_{A'}+\sum_{\tilde q \neq q} U_{q\tilde q}\ket{\cdot} \ket{\tilde
q}_{A'}$$ and is extended to a unitary matrix on the other states $\ket{q}_{A'}$. Observe that

\begin{align*}{\rm E}_{(q,q')}\left[\| \big(O_U\otimes O_U - (U\otimes \Id_{A'})\otimes (U\otimes
\Id_{B'}) \big)\ket{q,0}_A\ket{q',0}_B\ket{\Psi'} \ket{0}_{A'}\ket{0}_{B'}\|^2\right] \\
= {\rm E}_{(q,q')}\left[2 \sum_{(\tilde q,\tilde q')\neq (q,q')}\| U_{q\tilde q}\otimes U_{q'\tilde
q'}\ket{0}_A\ket{0}_B\ket{\Psi'}\|^2\right]\leq 4\eps,
\end{align*}
again for both when $(q,q')$ is sampled according to  $\pi(q,q')$  or according to $\pi(q)\pi(q')$.
 This means that for purposes of analysis we can replace Alice and Bob's transformation $U$ by
$O_U$, thereby replacing the transformation $U \otimes U$ on the message registers and $\ket{\Psi}$ by the
transformation $O_U \otimes O_U$ on the message space and $\ket{\tilde \Psi}=\ket{
\Psi'}\ket{0}_{A'}\ket{0}_{B'}$, at the expense of an error $4 \eps$ in statistical distance on the answer
probabilities of the classical test and the outcome probabilities in the quantum test. Since $O_U$ is
block-diagonal, the remainder of this claim follows exactly as in the proof of Claim \ref{claim:classical}.
\end{proof}

The SWAP-test now allows us to establish a set of inequalities which capture the ``almost commuting" property
of the operators. In what follows we will repeatedly use the following easy to verify fact.
\begin{fact}\label{fact:sumout}
Let $W^1,\ldots,W^k$ be  projectors such that $\sum_i W^i  = \Id$. Then $\sum_i \|W^i \ket{\Psi}\|^2 =
\|\ket{\Psi}\|^2$ for any vector $\ket{\Psi}$.
\end{fact}

\begin{claim}\label{claim:commute}
\begin{subequations}\label{eq:commute}
    \begin{gather}
\sum_{i,j=1}^{|Q|} \pi(q_i)\pi(q_j)\sum_{a_i,a'_j}\|(W_{q_i}^{a_i} \otimes W_{q_j}^{a'_j} - W_{q_j}^{a'_j}
 \otimes W_{q_i}^{a_i}) \ket{\Psi}\|^2 \leq
24\eps,\label{eq:commute1}\\
\sum_{i=1}^{|Q|}\pi(q_i)\sum_{a_i} \|(W_{q_i}^{a_i} \otimes \Id  -  \Id \otimes W_{q_i}^{a_i}) \ket{\Psi}\|^2
\leq 9\cdot 24 \cdot \eps.\label{eq:commute2}
\end{gather}
\end{subequations}
\end{claim}

\begin{proof}
As in the proof of Claim \ref{lemma:commute1}, the left-hand side of \eqref{eq:commute1} is four times the
probability to measure the first qubit in ``$-$" in the quantum test. For \eqref{eq:commute2}, using
Fact~\ref{fact:sumout}, for any fixed $q_j$ the following holds
\begin{align*}
\|(W_{q_i}^{a_i} \otimes \Id - \Id  \otimes W_{q_i}^{a_i}) \ket{\Psi}\|^2 & = \sum_{a'_j,a''_j} \|(W_{q_j}^{a'_j}W_{q_i}^{a_i} \otimes W_{q_j}^{a''_j} - W_{q_j}^{a'_j}
 \otimes W_{q_j}^{a''_j}W_{q_i}^{a_i}) \ket{\Psi}\|^2\\
&\leq  \sum_{a'_j,a''_j}  \Big( \|(W_{q_j}^{a'_j}W_{q_i}^{a_i} \otimes W_{q_j}^{a''_j} -W_{q_j}^{a'_j} W_{q_j}^{a''_j}\otimes W_{q_i}^{a_i} )\ket{\Psi} \| \\
& \qquad + \|(W_{q_j}^{a'_j} W_{q_j}^{a''_j}\otimes W_{q_i}^{a_i} - W_{q_i}^{a_i} \otimes W_{q_j}^{a''_j} W_{q_j}^{a'_j}) \ket{\Psi} \| \\
& \qquad  + \|(W_{q_i}^{a_i} \otimes W_{q_j}^{a''_j} W_{q_j}^{a'_j} - W_{q_j}^{a'_j}  \otimes
W_{q_j}^{a''_j}W_{q_i}^{a_i}) \ket{\Psi}\|\Big)^2.
 \end{align*}
\noindent We can bound the square of the sum of the three norms by $3$ times the sum of the norms squared,
and summing over $a_i$, averaging over $q_i,q_j$, and using $W_q^a W_{q}^{a'} = \delta_{a,a'} W_q^a$ for the
second norm and  Fact~\ref{fact:sumout} for the two others, we get three terms that are each bounded using
(\ref{eq:commute1}), concluding the proof of (\ref{eq:commute2}).
\end{proof}

\paragraph{Rounding to a classical strategy:} Order the questions in $Q$ such that $\pi(q_1) \geq \pi(q_2) \geq \ldots \geq
\pi(q_n)$. Define a joint distribution on answers $a_1,\ldots ,a_n$ as
$$D(a_1,\ldots ,a_n)=\|
(W_{q_n}^{a_n}\cdots W_{q_1}^{a_1} \otimes  \Id) \ket{\Psi}\|^2.$$ Fact \ref{fact:sumout} shows that $D$ is a
probability distribution, $\sum_{a_1,\ldots ,a_n}D(a_1,\ldots ,a_n) =1$.

We can interpret the distribution $D$ as follows: Before the game starts, the provers produce a joint list of
answers $a_1,\ldots , a_n$ as follows: They take the first part of $\ket \Psi$ and perform the projective
measurement corresponding to question $q_1$. They obtain an outcome $a_1$, which they record. They then take
the post-measurement state and perform on it the measurement corresponding to question $q_2$, and so on, each
time using the post-measurement state of one measurement as the input state of the next measurement. The
probability that the provers record answers $a_1,\ldots ,a_n$ is precisely $D(a_1,\ldots ,a_n)$.

Obviously neither quantum states nor measurements are needed to implement this constructed classical
strategy. Before the game starts, the provers simply compute $D$ for all inputs and sample from $D$ using
their shared randomness. When presented with questions $q_i,q_{j}$ they give the answer $a_i,a_{j}$, ignoring
all other answers in their sample. Hence the probability to answer $a_i,a_{j}$ in this case is given by the
marginal of $D$ with respect to $a_i$ and $a_j$, which we denote by $p_{class}(a_ia_{j}|q_iq_{j})$.

\begin{lemma}\label{lemma:2}
The (weighted) statistical distance between $p_{class}$ and $p_q$ is
$$\Delta(p_{class},p_q)=\sum_{q,q'}\pi(q,q')\sum_{a,a'}|p_{class}(a,a'|q,q')-p_q(a,a'|q,q')|
\leq 70 \cdot |Q|\cdot \eps^{1/4}.$$
\end{lemma}
Let us first show how this proves Lemma \ref{lemma:mainQMIP}. Since the quantum strategy of the provers
passes the classical test with probability at least $1-6\eps$, this means that the classical strategy wins
the original game with probability at least $1-6\eps-\Delta(p_{class},p_q)$ (where  $\Delta$ is the
dominating term), which we want to be larger than $s$. This is achieved for $\eps=\frac{s_q}{ |Q|^4}$ for a
sufficiently small constant $s_q$.
\end{proof}

\begin{proof}[Proof of Lemma \ref{lemma:2}.]

Let $q_{i},q_{j}$ be two questions. For convenience, let us introduce the notation $\sum_{{\bf a}}$ to denote
summing over $a_1,\ldots ,a_n$ and $\sum_{{\bf a}_{\neg i,j}}$ to denote summing over all $a_1,\ldots ,a_n$
except $a_i$ and $a_j$. Then the probability of answering $(a_{i},a_{j})$ to $(q_{i},q_{j})$  is
$p_{class}(a_{i}a_{j}|q_{i}q_{j}) = \sum_{{\bf a}_{\neg i,j}} \| (W_{q_n}^{a_n}\cdots W_{q_1}^{a_1} \otimes
\Id) \ket{\Psi} \|^2$ in the classical strategy, and
$p_{q}(a_{i},a_{j}|q_{i},q_{j})=\|W_{q_{i}}^{a_{i}}\otimes W_{q_{j}}^{a_{j}}\ket{\Psi}\|^2$ in the quantum
strategy. We wish to bound
$$\sum_{a_i,a_j}\big|p_{class}(a_{i}a_{j}|q_{i}q_{j})-p_{q}(a_{i},a_{j}|q_{i},q_{j})\big|   =
\sum_{a_i,a_j}\big| \sum_{{\bf a}_{\neg i,j}}
\| (W_{q_n}^{a_n}\cdots W_{q_1}^{a_1} \otimes \Id) \ket{\Psi} \|^2-\|W_{q_{i}}^{a_{i}}\otimes
W_{q_{j}}^{a_{j}}\ket{\Psi}\|^2\big|.$$  We now use a hybrid argument to go from the classical to the quantum
probability. The point is to eliminate the excess $W_{q}^{a}$ in $p_{class}$ with the help of Fact
\ref{fact:sumout}, which allows to eliminate a sum over $a$ that involves a $W_{q}^{a}$ on the {\em left}
side of all other operators in $\|\cdot\|^2$. To get all unwanted $W_q^a$ to be on the left, we move matrices
from one register to the other whenever they are on the {\em right}, closest to $\ket{\Psi}$, at the expense
of some error which we can bound using Eqs.(\ref{eq:commute}). More precisely we use the triangle inequality
for matrices $A,W,B,W'$
\begin{equation}\label{eq:triangle}
 \big|\|(AW \otimes BW')\ket{\Psi}\|-\|(AW' \otimes BW)\ket{\Psi}\|\big|\leq
\|(A \otimes B)[W \otimes W'-W'\otimes W]\ket{\Psi}\|,
\end{equation}
where $A$ and $B$ will be sequences of $W_q^a$ and $W$ or $W'$ are either one of the $W_q^a$ or the identity.

To describe the sequence along which  we move the matrices around, let us use the shorthand notation $W_k$
for $W_{q_k}^{a_k}$. At each step we will interchange either $W_k \otimes \Id \leftrightarrow \Id \otimes
W_k$ or $W_i \otimes W_k \leftrightarrow W_k \otimes W_i$ whenever they are on the right. If $i
> j$ we proceed according to the sequence
 \begin{align*}
 & W_n\cdots W_1\otimes \Id  \rightarrow W_n\cdots W_2\otimes W_1  \rightarrow  W_n\cdots W_3\otimes W_1W_2
\rightarrow \cdots \rightarrow  W_n\cdots W_{i+1}W_i\otimes W_1\cdots W_{i-1} \\
 & \rightarrow W_n\cdots W_{i+1}W_{i-1}\otimes W_1\cdots W_{i-2}W_{i}  \rightarrow   W_n\cdots W_{i+1}W_{i-1}W_i\otimes
W_1\cdots W_{i-2} \\ & \rightarrow W_n\cdots W_{i+1}W_{i-1}W_{i-2}\otimes W_1\cdots W_{i-3}W_i \rightarrow
\cdots \rightarrow W_n\cdots W_{i+1}W_{i-1}\cdots W_{j+1}W_i\otimes W_1\cdots W_j.
\end{align*}
Note that the last term in the sequence, when summed over ${\bf a}_{\neg i,j}$, equals $p_q(a_ia_j|q_iq_j)$
because of Fact \ref{fact:sumout}, i.e. $\sum_{{\bf a}_{\neg i,j}}\|W_n\cdots W_{j+1}W_i\otimes W_1\cdots
W_j\ket{\Psi}\|^2=\|W_i\otimes W_j \ket{\Psi}\|^2=p_q(a_ia_j|q_iq_j)$. Now we can write a telescopic sum
according to this sequence as
 \begin{align*}
\sum_{a_i,a_j}|p_{class}(a_{i}a_{j}|q_{i}q_{j}) - p_q&(a_{i}a_{j}|q_{i}q_{j})|=\sum_{a_i,a_j}\Big|\sum_{{\bf
a}_{\neg i,j}}\|W_n\cdots W_1\otimes \Id \ket{\Psi}\|^2
-\sum_{{\bf a}_{\neg i,j}}\|W_n\cdots W_2\otimes W_1 \ket{\Psi}\|^2\\
&  +\sum_{{\bf a}_{\neg i,j}}\|W_n\cdots W_2\otimes W_1 \ket{\Psi}\|^2-\sum_{{\bf a}_{\neg i,j}}\|W_n\cdots
W_3\otimes
W_1W_2 \ket{\Psi}\|^2 +\cdots \Big|\\
&\leq \sum_{{\bf a}}\big|\|W_n\cdots W_1\otimes \Id \ket{\Psi}\|^2 -\|W_n\cdots W_2\otimes W_1
\ket{\Psi}\|^2\big|+\sum_{{\bf a}}\big|\cdots \big|+\cdots ,
 \end{align*}
where we used the triangle inequality. Using $|a^2-b^2|=|a-b|\cdot|a+b|$, and the triangle inequality as in
(\ref{eq:triangle}), the first term is bounded by
\begin{align*}
&\sum_{{\bf a}} \|W_n\cdots W_2 [W_1\otimes \Id-\Id\otimes W_1]\ket{\Psi}\|\cdot (\|W_n\cdots W_1\otimes \Id
\ket{\Psi}\| + \|W_n\cdots W_2\otimes W_1
\ket{\Psi}\|)\\
& \leq \sqrt{\sum_{{\bf a}} \|W_n\cdots W_2 [W_1\otimes \Id -\Id\otimes W_1]\ket{\Psi}\|^2}\sqrt{\sum_{{\bf
a}} (\|W_n\cdots W_1\otimes \Id \ket{\Psi}\| + \|W_n\cdots W_2\otimes W_1 \ket{\Psi}\|)^2},
\end{align*}
where we used Cauchy-Schwarz for the inequality. We obtain similar expressions for all other terms.  We can
bound the second square root by $\sqrt{2+2}=2$, using $(a+b)^2 \leq 2a^2+2b^2$ and Fact \ref{fact:sumout}.
Assembling all the  terms, and using Fact \ref{fact:sumout} to eliminate all the matrices to the left of the
square brackets, we obtain
\begin{align}\label{eq:threeterms}
\sum_{a_i,a_j}|p_{class}(a_{i}a_{j}|q_{i}q_{j}) - p_q(a_{i}a_{j}|q_{i}q_{j})| \leq &\;2 \sum_{i'=1}^{i-1}
\sqrt{\sum_{a_{i'}}
\|[W_{i'}\otimes \Id-\Id\otimes W_{i'}]\ket{\Psi}\|^2} \nonumber \\
+&\;2(|i-j|+1)\sqrt{\sum_{a_i} \|[\Id \otimes W_i- W_i \otimes \Id ] \ket{\Psi}\|^2}\nonumber \\
+ & \;2 \sum_{i'=j+1}^{i-1}\sqrt{\sum_{a_i,a_{i'}}\|[W_{i}\otimes W_{i'}-W_{i'}\otimes W_{i}]\ket{\Psi}\|^2}.
\end{align}
For $j>i$ we obtain exactly the same sequence and the same bounds in Eq. (\ref{eq:threeterms}) with $i$ and
$j$ interchanged. The only difference is that now the last term in the sequence, when summed over ${\bf
a}_{\neg i,j}$ gives $\|W_j \otimes W_i\ket{\Psi}\|^2$, so we need to use symmetry of $\ket{\Psi}$ to
conclude that this equals to $\|W_i \otimes W_j\ket{\Psi}\|^2$. For $i=j$  we follow the sequence until
$W_n\cdots W_{i+1}W_i\otimes W_{1}\cdots W_{i-1}$ and then use $W_i=W_i^2$ to continue as $W_n\cdots
W_{i+1}W_iW_i\otimes W_{1}\cdots W_{i-1} \rightarrow W_n\cdots W_{i}\otimes W_{1}\cdots W_{i-1} W_i$, so we
just get the first term in Eq. (\ref{eq:threeterms}), but summed until $i$.

Now $\Delta(p_{class},p_q)$ is bounded by the average over $(q_{i},q_{j})$ picked according to the
distribution $\pi$ of the sum of the three terms appearing in (\ref{eq:threeterms}). We show how to bound
each of them. For the first term
\begin{align*}
&2 \sum_{i,j=1}^{|Q|} \pi(q_{i},q_{j}) \sum_{i'=1}^{i}\sqrt{\sum_{a_{i'}} \|(W_{q_{i'}}^{a_{i'}}\otimes \Id -
\Id \otimes W_{q_{i'}}^{a_{i'}}) \ket{\Psi}\|^2} \\
&=2 \sum_{i=1}^{|Q|} \pi(q_{i}) \sum_{i'=1}^{i}\sqrt{\sum_{a_{i'}} \|(W_{q_{i'}}^{a_{i'}}\otimes \Id -
\Id \otimes W_{q_{i'}}^{a_{i'}}) \ket{\Psi}\|^2} \\
&\leq  2\sum_{i=1}^{|Q|} \sum_{i'=1}^{|Q|}\pi(q_{i'})\sqrt{\sum_{a_{i'}} \|(W_{q_{i'}}^{a_{i'}}\otimes \Id
- \Id \otimes W_{q_{i'}}^{a_{i'}}) \ket{\Psi}\|^2}\\
&\leq  2 |Q| \big(\sum_{i'=1}^{|Q|}\pi(q_{i'})\sum_{a_{i'}} \|(W_{q_{i'}}^{a_{i'}}\otimes \Id - \Id \otimes
W_{q_{i'}}^{a_{i'}}) \ket{\Psi}\|^2\big)^{1/2} \leq 2|Q|\sqrt{9 \cdot 24\eps},
 \end{align*}
where the first equality uses the fact that the inner sum does not depend on $j$, the second inequality uses
$\pi(q_{i}) \leq \pi(q_{i'})$, the third inequality uses the fact that the square of the expectation is not
greater than the expectation of the square, and the last inequality uses Eq. (\ref{eq:commute2}). The second
term can be bounded in a similar fashion
\begin{align*}
&2\sum_{i,j=1}^{|Q|} \pi(q_{i},q_{j}) (|i-j|+1)\sqrt{ \sum_{a_{i}}\|(\Id \otimes W_{q_{i}}^{a_{i}} -
W_{q_{i}}^{a_{i}} \otimes \Id)\ket{\Psi}\|^2}\\&\leq
 2 |Q| \sum_{i=1}^{|Q|} \pi(q_{i}) \sqrt{\sum_{a_{i}}\|(\Id \otimes W_{q_{i}}^{a_{i}} - W_{q_{i}}^{a_{i}}
 \otimes \Id)\ket{\Psi}\|^2} \leq 2|Q|\sqrt{9 \cdot 24\eps}.
\end{align*}
Finally the last term, using again that the inner sum does not depend on $j$, that the square of the
expectation is bounded by the expectation of the square and Cauchy-Schwarz for the sum over $i'$, can be
bounded by
\begin{align}\label{eq:thirdterm}
&2\sum_{i=1}^{|Q|} \pi(q_{i})  \sum_{i'=1}^{i-1}\sqrt{\sum_{a_{i},a_{i'}} \|(W_{q_{i}}^{a_{i}} \otimes
W_{q_{i'}}^{a_{i'}} - W_{q_{i'}}^{a_{i'}} \otimes W_{q_{i}}^{a_{i}}) \ket{\Psi}\|^2} \nonumber\\ & \leq
2\Big(\sum_{i=1}^{|Q|} \pi(q_{i}) \Big(\sum_{i'=1}^{i-1}\sqrt{\sum_{a_{i},a_{i'}} \|(W_{q_{i}}^{a_{i}}
\otimes W_{q_{i'}}^{a_{i'}} - W_{q_{i'}}^{a_{i'}} \otimes W_{q_{i}}^{a_{i}})
\ket{\Psi}\|^2}\;\;\Big)^2\Big)^{1/2}\nonumber \\
& \leq  2\sqrt{|Q|}  \Big(\sum_{i=1}^{|Q|} \pi(q_{i}) \sum_{i'=1}^{i-1}\sum_{a_{i},a_{i'}}
\|(W_{q_{i}}^{a_{i}} \otimes W_{q_{i'}}^{a_{i'}} - W_{q_{i'}}^{a_{i'}} \otimes W_{q_{i}}^{a_{i}})
\ket{\Psi}\|^2\Big)^{1/2}.
\end{align}
We decompose the sum inside the square root in the last line of (\ref{eq:thirdterm}) into two parts with
$\pi(q_{i})\geq 1/h$ and $\pi(q_{i})< 1/h$ (with $h$ to be determined later). If $\pi(q_{i})\geq 1/h$, then
$\pi(q_{i'})\geq 1/h$ for $i'\leq i$ so $1\leq h\pi(q_{i'})$. Therefore, using (\ref{eq:commute1}), the term
in parenthesis in (\ref{eq:thirdterm}) is bounded by
\begin{align*}
 &\sum_{i:\pi(q_{i})\geq 1/h} \sum_{i'=1}^{i-1} h\pi(q_{i'})\pi(q_{i})\sum_{a_{i},a_{i'}} \|(W_{q_{i}}^{a_{i}}
 \otimes W_{q_{i'}}^{a_{i'}} - W_{q_{i'}}^{a_{i'}}
\otimes W_{q_{i}}^{a_{i}}) \ket{\Psi}\|^2 \\
&+ \frac{1}{h} \sum_{i:\pi(q_{i})\leq 1/h}\sum_{i'=1}^{i-1}\sum_{a_{i},a_{i'}} \|(W_{q_{i}}^{a_{i}} \otimes
W_{q_{i'}}^{a_{i'}} - W_{q_{i'}}^{a_{i'}} \otimes W_{q_{i}}^{a_{i}}) \ket{\Psi}\|^2\leq  24h \eps + 4|Q|^2/h,
\end{align*}
where we have bounded the first part using (\ref{eq:commute1}) and the second part, using triangle inequality
and Fact \ref{fact:sumout}
$$\sum_{a_{i},a_{i'}} \|(W_{q_{i}}^{a_{i}} \otimes W_{q_{i'}}^{a_{i'}} - W_{q_{i'}}^{a_{i'}} \otimes
W_{q_{i}}^{a_{i}}) \ket{\Psi}\|^2\leq \sum_{a_{i},a_{i'}} (\|(W_{q_{i}}^{a_{i}} \otimes
W_{q_{i'}}^{a_{i'}})\ket{\Psi}\|+\|(W_{q_{i'}}^{a_{i'}} \otimes W_{q_{i}}^{a_{i}}) \ket{\Psi}\|)^2 \leq 4.$$
The optimal $h$ is $|Q| / \sqrt{6\eps}$, which gives a bound of $4\cdot 24^{1/4}  |Q|  {\eps}^{1/4}$ for the
third (dominant) term in $\Delta(p_{class},p_q)$ (after taking the square root). Hence $\Delta(p_{class},p_q)
\leq 70 |Q| {\eps}^{1/4}$.
\end{proof}

\section{Hardness of three-prover entangled classical games}\label{sec:MIP}

In this section we prove Theorem~\ref{thm:main} for three-prover entangled classical games, which we now state as:

\begin{theorem}\label{thm:MIPmain}
There is a constant $s_3>0$ such that it is \NP-hard to decide, given an entangled three-prover classical
game with a constant number of answers, whether its value is $1$ or less than $1-\eps$ for
$\eps=\frac{s_3}{|Q|^2}$.
\end{theorem}

As in the case of quantum games, we will prove this by a reduction from the PCP Theorem. This time, however,
we will essentially preserve the number of answers in the modified game.

We begin by describing how to modify any two-prover classical game $G(2,\pi,V)$ (which is assumed to be
symmetric per Lemma~\ref{lem:symmetry1}) to a three-prover classical game $G'$ of equal or higher value.

\paragraph{The modified three-prover game.}
In the constructed game $G'$ the verifier chooses one of the provers uniformly at random. Rename the chosen
prover Alice and call the other provers Bob and Cleve. The verifier samples questions $q$ and $q'$ according
to $\pi(q,q')$. He sends question $q$ to Alice, and question $q'$ to both Bob and Cleve. He receives answers
$a$, $a'$, and $a''$, respectively, and accepts iff the following are both true:

\smallskip
\noindent{\bf Classical Test:} The answers of Alice and Bob would win the game $G$, i.e., $V(aa'|qq') = 1$.

\smallskip
\noindent{\bf  Consistency:} Bob and Cleve give the same answer, i.e., $a' = a''$.

\smallskip
{\em Remarks:} Note that unlike the quantum case, the verifier performs both tests at the same time.  The
consistency test plays the role of the SWAP test, limiting the advantage gained by sharing entanglement.

Again it is clear that the value of the constructed game is at least as large as the value of the original
game $G$: if the provers reply according to an optimal classical strategy (which can be assumed to be
symmetric per Lemma \ref{lem:symmetry1}) they always pass the consistency-test. Also, it is clear in this
case that the size of the description of the constructed game is linearly related to the size of the
description of the original game, hence we have the same complexity parameter.

To prove Theorem \ref{thm:MIPmain}, we need to show the following.

\begin{lemma}\label{lemma:mainMIP}
If $\omega^*(G')>1-\eps$ then $\omega(G)> s$.
\end{lemma}
\begin{proof}
Consider a quantum strategy for $G'$ that succeeds with probability $1-\eps$.\footnote{Again, as in Section
\ref{sec:QMIPapprox}, we in fact consider a strategy with finite entanglement that has success probability
$1-\eps-\delta$ for some  $\delta=O(\eps)$, which we will not write.} Since the game $G'$ is symmetric, we
can assume that this strategy is symmetric, per Lemma~\ref{lem:symmetry}. Suppose that the provers share a
symmetric state $\ket \Psi \in {\cal H}^{\otimes 3}$.  Let $\rho^\text{AB} = \tr_{{\cal H}_3} \ket \Psi \bra
\Psi$ be the reduced density matrix of $\ket \Psi\bra \Psi$ on Alice and Bob. When asked question $q_i$, each
prover measures their part of $\ket \Psi$. Following standard arguments (extending the private space of the
provers) we can assume that this measurement is projective. Let $W^{a_i}_{q_i}$ be the projector
corresponding to question $q_i$ and answer $a_i$. This defines the quantum strategy for
  $G'$; it passes the classical test with probability
$$   \pi_1 = \sum_{aa'qq'} \pi(q,q') V(aa'|qq') p_q(aa'|qq'),
$$
where
\begin{align}\label{eq:4}
  p_q(aa'|qq') =  \tr \left(W^{a}_{q} \otimes W^{a'}_{q'} \rho^\text{AB}\right) = \bra \Psi W^{a}_{q} \otimes
  W^{a'}_{q'} \otimes \Id \ket \Psi.
\end{align}
It passes the consistency test with probability $
  \pi_2 = \sum_{q} \pi(q) \pi_2(q),
$ where $\pi(q)$ is the marginal of $\pi(q,q')$ and
\begin{align}
\label{eq:7}
  \pi_2(q) = \sum_a \tr \left(W^{a}_{q} \otimes W^{a}_{q} \rho^\text{AB}\right) = \sum_a \bra \Psi W^{a}_{q} \otimes W^{a}_{q}
  \otimes \Id \ket \Psi,
\end{align}
where we made use of the symmetry.  Note that $\pi_1, \pi_2 \geq 1 - \eps$.

Eqs.~(\ref{eq:4}) and (\ref{eq:7}) clarify the role of the third prover, Cleve.  His main purpose is
\emph{not} to allow the two tests to be performed at the same time: Indeed, it is possible to modify the
protocol so that the verifier chooses two of the provers at random (say Alice and Bob) and only sends
questions to them, not interacting with the third prover at all.\footnote{With probability $p$, he sends them
different questions and performs the classical test; with probability $1-p$, he sends the same question and
performs the consistency test---this modification does not materially change our conclusions, but it does
weaken the bounds in Theorem~\ref{thm:MIPmain}.}  Cleve's presence would not be important if the provers were
executing a classical strategy, but it can (and does) make a difference if their strategy requires
entanglement. Indeed, if there were only two provers, then they could share any state $\rho^\text{AB}$,
whereas here we require that $\rho^\text{AB}$
 be \emph{extendable}, i.e., it must be the reduced density
matrix of a symmetric tripartite state.  To give a concrete example,
it is not possible for $\rho^\text{AB}$ to be the maximally entangled
state $\ket {\Psi^-} \bra {\Psi^-}$.  This is termed \emph{monogamy of
  entanglement}~\cite{werner89:_applic_inequal_quant_state_exten_probl}.

\paragraph{Rounding to a classical strategy:} We construct a classical strategy for $G$ from the quantum strategy for $G'$ in a similar
 fashion as in the  case of quantum games, with
\begin{align}\label{eq:2}D(a_1,\ldots ,a_n,a'_1,\ldots ,a'_n)= \| W_{q_n}^{a_n}\cdots W_{q_1}^{a_1} \otimes
W_{q_n}^{a'_n} \cdots W_{q_1}^{a'_1}\otimes \Id \ket \Psi\|^2.
\end{align}
\noindent where $q_1,\ldots,q_n$ is an ordering of the questions in $Q$ such that $\pi(q_1)\geq
\pi(q_2)\geq\ldots\geq \pi(q_n)$.\footnote{Note that $D$ differs slightly from Sec. \ref{sec:QMIPapprox}.
Here each prover gets a separate list of answers. This form is more convenient here.} As before, we define
$p_{class}(a_i,a'_j|q_i,q_j)$ to be the marginal of $D$ on $a_i,a'_j$. The structure of our proof that this
strategy is a good one is very similar to the quantum case. The details, however, are a little different.

\begin{lemma}\label{lemma:3}
The (weighted) statistical distance between $p_{class}$ and $p_q$ is
$$\Delta(p_{class},p_q)=\sum_{q,q'}\pi(q,q')\sum_{a,a'}|p_{class}(a,a'|q,q')-p_q(a,a'|q,q')| \leq 12 |Q|
\sqrt{\eps}.$$
\end{lemma}
We first show how this Lemma proves Lemma \ref{lemma:mainMIP}. Since the strategy in the entangled game
passes the classical test with probability at least $1-\eps$, the classical strategy succeeds in the original
game with probability at least $1- \eps - \Delta \geq 1-\eps-12 |Q|\sqrt{\eps}$. For $\eps=\frac{s_3}{|Q|^2}$
 for sufficiently small constant $s_3$, this probability is larger than $s$.
\end{proof}
This Lemma is the corresponding version of Lemma~\ref{lemma:2}.  Why is it true?  Rather than showing that
the order of measurements is not important as we did in the quantum case (although it will turn out in
hindsight that this is true), we show that each measurement does not disturb $\rho^\text{AB}$ very much.  The
key observation is as follows.  Assume the provers pass the consistency test with high probability.  If a
particular measurement result occurs with certainty, the quantum state cannot be changed by the measurement.
We use this fact in the following way: suppose Cleve were to perform the measurement corresponding to
question $q$ and assume he obtains an outcome $a$. Then, if Bob is asked question $q$, he must also give
answer $a$ with high probability.  So his measurement does not change the quantum state much.  But, since
quantum theory is no-signalling, it cannot matter who measured first.  It follows that Bob's measurement does
not change $\rho^\text{AB}$ much.  Note that only the bipartite state $\rho^\text{AB}$ is approximately
unchanged---Bob's measurement can change the tri-parite state $\ket \Psi \bra \Psi$ considerably.  We then
use a hybrid argument to show that performing all the measurements one after the other also leaves
$\rho^\text{AB}$ approximately unchanged.  This part of the proof mirrors the proof of Lemma~\ref{lemma:2}.

\begin{proof}[Proof of Lemma~\ref{lemma:3}]  Let ${\cal W}_q$ be the superoperator corresponding to the
projective measurement $q$, i.e., ${\cal W}_q(\sigma) := \sum_a W^a_q \sigma (W^a_q)^\dagger$ is the
post-measurement state after performing $\{W^a_q\}$ on state $\sigma$.

To quantify how much a measurement changes a state we use Winter's gentle measurement lemma.

\begin{lemma}[Lemma~I.4 \cite{winter:thesis}]\label{lem:winter}
Let $\rho$ be a state and $X$ be a positive matrix with $X\leq \Id$ and $0\leq \Tr\rho X$. Then,
$$ \left\|\rho - \sqrt{X}\rho\sqrt{X}\right\|_1\leq 3\sqrt{1- \Tr X\rho}.$$
\end{lemma}

The following simple corollary quantifies how much the measurement ${\cal W}_q \otimes \Id$ changes
$\rho^\text{AB}$:
\begin{claim}\label{claim:winter}The trace distance between ${\cal W}_q \otimes \Id (\rho^\text{AB})$ and
$\rho^\text{AB}$ is bounded by
  \begin{align*}
    \| {\cal W}_q \otimes \Id (\rho^\text{AB}) - \rho^\text{AB} \|_1 \leq 6 \sqrt{1 - \pi_2 (q)}.
  \end{align*}
\end{claim}
\begin{proof}Using ${\cal W}_q \otimes \Id (\rho^\text{AB}) = \tr_{{\cal H}_3}({\cal W}_q \otimes \Id \otimes \Id
(\ket \Psi\bra \Psi))$ and $\rho^\text{AB} = \tr_{{\cal H}_3}(\Id \otimes \Id \otimes {\cal W}_q (\ket \Psi\bra \Psi))$,
by monotonicity of the trace distance under partial trace,
  \begin{align*}
    \| {\cal W}_q \otimes \Id (\rho^\text{AB}) - \rho^\text{AB} \|_1 &\leq     \| {\cal W}_q \otimes \Id \otimes \Id (\ket \Psi\bra \Psi) - \Id \otimes \Id \otimes {\cal W}_q (\ket \Psi\bra \Psi) \|_1\\
&\leq \| {\cal W}_q \otimes \Id \otimes \Id (\ket \Psi\bra \Psi)- \sum_a W_q^a \otimes \Id \otimes W_q^a \ket \Psi\bra \Psi W_q^a \otimes \Id \otimes W_q^a\|_1 \nonumber\\
& \hspace{0.5cm} + \|\sum_a W_q^a \otimes \Id \otimes W_q^a \ket \Psi\bra \Psi W_q^a \otimes \Id \otimes W_q^a  - \Id \otimes \Id \otimes {\cal W}_q (\ket \Psi\bra \Psi) \|_1\\
&\leq 2 \| \sum_a W_q^a \otimes \Id \otimes W_q^a \ket \Psi\bra \Psi W_q^a \otimes \Id \otimes W_q^a - \Id \otimes \Id \otimes {\cal W}_q (\ket \Psi\bra \Psi) \|_1\\
&\leq 6 \sqrt{1 - \pi_2(q)},
  \end{align*}
by the triangle inequality, symmetry, and then taking $\rho = \bigoplus_a  W_q^a \otimes \Id \otimes \Id \ket
\Psi \bra \Psi W_q^a \otimes \Id \otimes \Id $ and $X = \bigoplus_a \Id \otimes \Id \otimes W_q^a $ in Lemma~\ref{lem:winter}.
\end{proof}

For $1\leq i, j \leq n$, let
$$
    \rho^\text{AB}(i,j) := ({\cal W}_{q_{i-1}}\circ \cdots \circ {\cal W}_{q_{1}}) \otimes ({\cal W}_{q_{j-1}} \circ
    \cdots \circ
    {\cal W}_{q_{1}})\rho^\text{AB}
$$
Then
$$
  p_{class}(a_ia_j'|q_i q_j') = \tr \left((W^{a_i}_{q_i} \otimes W^{a_j'}_{q'_j}) \rho(i,j)\right)
$$
Hence if we can bound $\|\rho(i,j) - \rho\|_1$, then we can bound $\sum_{a_i, a'_j} |p_{class}(a_i a_j'|q_i
q_j') - p_q(a_i a_j'|q_i q_j')|$, since the trace distance between two states is an upper bound on the
variation distance of the probability distribution resulting from making any measurement on those two states.

The following technique was introduced by Ambainis, Nayak, Ta-Shma, and U. Vazirani~\cite{ambainis:rac02} and
has been used extensively by Aaronson~\cite{aaronson:05,aaronson06:qmaqpoly}.
\begin{claim}  The trace distance between $\rho^\text{AB}(i,j)$ and $\rho^\text{AB}$ is bounded by
 $$
\| \rho^\text{AB}(i,j) - \rho^\text{AB}\|_1 \leq 6\sum_{i' = 1}^{i-1} \sqrt{1-\pi_2(q_{i'})} + 6\sum_{j' =
1}^{j-1} \sqrt{1-\pi_2(q_{j'})}.
 $$
\end{claim}
\begin{proof} Proof by induction.  The claim is clearly true for $(i,j) = (1,1)$.  Given it is true for a particular value of $(i,j)$, we show it is also true for $(i+1,j)$.  In view of the symmetry, this is sufficient to establish the claim. We have
  \begin{align*}
\| \rho^\text{AB}(i+1,j) - \rho^\text{AB}\|_1 &\leq \| \rho^\text{AB}(i+1,j) - {\cal W}_{q_{i}} \otimes \Id (\rho^\text{AB}) \|_1 + \| {\cal W}_{q_{i}} \otimes \Id (\rho^\text{AB})- \rho^\text{AB}\|_1\\
&\leq \| {\cal W}_{q_i} \otimes \Id \left( \rho^\text{AB}(i,j) - \rho^\text{AB} \right) \|_1 + 6\sqrt{1 - \pi_2(q_i)}\\
&\leq \| \rho^\text{AB}(i,j) - \rho^\text{AB} \|_1 + 6\sqrt{1 - \pi_2(q_i)},
\end{align*}
where we used the triangle inequality, Claim~\ref{claim:winter}, and monotonicity of the trace distance.
\end{proof}

Putting everything together, it follows that
\begin{align*}
\Delta(p_{class},p_q)&\leq \sum_{i,j=1}^n\pi(q_i,q'_j) \| \rho^\text{AB} (i,j) - \rho^\text{AB} \|_1 \\
& \leq 6 \sum_{i,j=1}^n\pi(q_i,q'_j) \left(\sum_{i' = 1}^{i-1} \sqrt{1-\pi_2(q_{i'})} + \sum_{j' = 1}^{j-1} \sqrt{1-\pi_2(q_{j'})}\right)\\
& \leq 12 \sum_{i=1}^n \sum_{i' = 1}^{i-1} \pi(q_i) \sqrt{1-\pi_2(q_{i'})}\\
& \leq 12 |Q| \sum_{i' = 1}^{n} \pi(q_{i'})\sqrt{1-\pi_2(q_{i'})}\\
& \leq 12 |Q|\sqrt{1- \pi_2} \leq 12 |Q| \sqrt{\epsilon},
\end{align*}
since $\pi_2 = \sum_{q} \pi(q) \pi_2(q) \geq 1-\eps$ and $\sqrt{1-x}$ is concave.
\end{proof}

\section{Hardness for two-prover classical entangled games}\label{sec:IP}

In this section we prove our main theorem for two-prover entangled classical games. It shows that it is
$\PSPACE$-hard to decide, given a succinct entangled two-prover classical game, whether its value is $1$ or
less than $1-\eps$ for $\eps=\frac{1}{\poly(|x|)}$. To state the result, we need some further definitions to
clarify the notion of succinctly given games and state the connection between \PSPACE\ and multi-round
single-prover games.

\begin{definition}
  \label{definition:1}
A language $L$ is in $\MIP^*_{c,s}(N,1)$ if, for all $x \in L$, there is a polynomial time (in $|x|$) mapping
from $x$ to classical one-round games $G_x(N,\pi_x, V_x)$, such that it is possible to sample from $\pi_x$ in
polynomial time and compute the predicate $V_x$ in polynomial time and
  \begin{itemize}
  \item Completeness: for all $x \in L$,  the entangled value $\omega^*(G_x) \geq c$, and
\item  Soundness: for all $x \not \in L$,  the entangled value $\omega^*(G_x)\leq s$.
  \end{itemize}
\end{definition}
Note that in this scenario the game is given {\em succinctly}: it is given by a description of $V$ (as a
polynomial time circuit, for instance, which implies that $|Q|,|A|=2^{\poly(|x|)}$) and a polynomial size
description of $\pi$, which can be sampled in polynomial time. Hence the complexity parameter here is $|x|$,
and $|Q|$ and $|A|$ are exponential.

We also require the notion of single-prover games with multiple rounds. We modify Definition  \ref{def:games}
to account for games with multiple rounds. Here we will only consider {\em non-adaptive} games: the
probability distribution on questions in $Q$ for each round $k$ does not depend  on the answers received in
previous rounds, which is sufficient for \PSPACE\ (see Theorem  \ref{thm:PSPACE}). However, we allow for the
possibility that the questions asked in each round depend on the questions asked in previous
rounds.\footnote{Note that this is equivalent to having a joint distribution on the questions, where we
obtain the distribution on the $i$th question as the corresponding marginal.} In other words a one-player
$r$-round game $G(1,\pi_r,V_r)$ is given by a joint distribution $\pi: Q^r \rightarrow [0,1]$, and a
predicate $V_r:A^r \times Q^r \rightarrow \{0,1\}$ (i.e. the verifier accepts or rejects as a function of all
the answers received in all rounds). The strategy is now a set of $r$ functions $W_k$, where the $k$th
function can depend on the previous questions and answers. The class \IP$(r)$ is given by Definition
\ref{definition:1} when the game is a single-prover multi-round game with $r$ rounds. We omit reference to
$r$ and write \IP\ when the number of rounds is polynomial in $|x|$.

\begin{theorem}\label{thm:PSPACE}\cite{Sha92}
There is a constant $s_{IP}\geq 0$ such that $\PSPACE=\IP_{1,s_{IP}}$. Moreover there are ``public-coin
non-adaptive" \IP-protocols for \PSPACE, i.e. such that in each round the distribution on the questions is
independent  of the answers of the prover and of  other rounds \cite{GolSip89RC,Shen92}.
\end{theorem}

With these notions in place we can state our main result for two-prover classical entangled games.

\begin{theorem}\label{thm:IPmain}
$\PSPACE \subseteq \MIP^*(2,1)_{1,1-\eps}$ for $\eps=\frac{1}{\poly(|x|)}$, where $|x|$ is the input size.
\end{theorem}

We note that if a parallel repetition theorem could be established for
classical two-prover entangled games, then the
containment in Theorem~\ref{thm:IPmain} could be improved to $\textsc{PSPACE} \subseteq
\MIP^*(2,1)_{1,s}$ with constant or even exponentially small $s$. This
is a particularly interesting direction to pursue, in light of the perfect
parallel repetition theorem for entangled XOR games of Cleve et
al.~\cite{Cleve:parallel} (which uses the SDP-description on the value
of these games).

To prove Theorem \ref{thm:IPmain} we use the \PSPACE-characterization in
Theorem \ref{thm:PSPACE} and show the following.
 \begin{lemma}\label{lemma:IP}
There is a constant $s_2 \geq 0$ such that for every succinctly given single-prover $r$-round non-adaptive
game $G(1,\pi_r,V_r)$, of value $\omega(G)$ with questions $Q$ and answers $A$, there is a two-prover
one-round classical game $G_c(2,\pi,V)$ with questions $Q^r$ and answers $A^r$ with entangled value
$\omega^*(G_c)\geq \omega(G)$ such that if $\omega^*(G_c)>1-\eps$ then $\omega(G)>s_{IP}$ for
$\eps=\frac{s_2}{r^2}$. Moreover, a succinct description of $G_c$ can be computed from a description of $G$
in polynomial time, and sampling $\pi$ and computing $V$ can be done in polynomial time.
 \end{lemma}

Lemma \ref{lemma:IP} shows $\IP(r)_{1,s_{IP}} \subseteq \MIP(2,1)^*_{1,1-\frac{s_2}{r^2}}$, and combined with
Theorem \ref{thm:PSPACE} gives Theorem \ref{thm:IPmain}.

The rest of this section is dedicated to the proof of Lemma \ref{lemma:IP}. It follows the main traits of the
proofs of the previous two hardness results. Our construction of the two-prover one-round game uses a
protocol of \cite{CaiConLip94JCSS} used to prove that $\PSPACE$ has two-prover one-round systems. We show
that this protocol remains sound even against entangled provers, albeit with larger soundness. To prove this
we again use the consistency test with the extra prover to extract almost commuting conditions on the
operators of the provers. This allows us to round in a similar fashion from a good strategy for the entangled
game to a strategy for the single prover game which succeeds with relatively large probability.

\paragraph{The modified two-prover game.}
 In the constructed game $G_c$, the verifier samples a series of questions $q_1,\ldots,q_r$
according to the distribution $\pi_r(q_1,\ldots ,q_r)$. He picks a $k$ uniformly at random in
$\{1,\ldots,r\}$, and sends questions $q_1,q_2, \ldots ,q_r$ to Alice and $q_1,q_2, \ldots, q_k$ to Bob. He
receives answers $a=a_1,\ldots,a_r$ from Alice and $a'=a'_1,\ldots,a'_k$ from Bob. He accepts if and only if
the following are both true:

\smallskip

{\bf Classical Test} The answers Alice gives would win the game $G$: $V(a_1\ldots a_n|q_1 \ldots q_n) = 1$.

\smallskip
{\bf Consistency Test} For all $i$ in $\{1,\ldots,k\}$, $a_i=a'_i$.
\smallskip

{\em Remark:} It is again obvious that the value of the new game is lower bounded by the value of the
original game: If both provers reply according to an optimal classical strategy, then they will always give
consistent answers, so their acceptance probability is exactly $\omega(G)$.

It is also easy to see that the constructed game has the same complexity as the original game. The new
verifier essentially implements the original verifier and the consistency test, which can be described in
linear time in $A^r$. The sampling procedure also has the same complexity as sampling from the original
$\pi_r$. And obviously it is possible to compute the new game from the original game in polynomial time.

To prove Lemma \ref{lemma:IP} we need to show the following.

\begin{lemma}\label{lemma:last}
If $\omega^*(G_c)>1-\eps$ then $\omega(G)>s_{IP}$.
\end{lemma}

\begin{proof}

Consider a quantum strategy for $G'$ that succeeds with probability $1-\eps$.\footnote{Again, as in Section
\ref{sec:QMIPapprox}, we in fact consider a strategy with finite entanglement that has success probability
$1-\eps-\delta$ for some  $\delta=O(\eps)$, which we will not write.} For any sequence of questions
$q_1,\ldots,q_r$ we define $\q{k}$ to be the sequence $q_1,\ldots,q_k$. Similarly, for any sequence
$a=a_1,\ldots,a_r$ of possible answers we will denote its prefix $a_1,\ldots,a_k$ by $\an{k}$. Note that when
we write $\an{k}$ and $\an{l}$ for some $1\leq k,l \leq r$ we refer to substrings of the {\em same} string
$a=a_1,\ldots,a_r$, whereas we will write $\an{k}$ and $\anp{l}$ if we refer to {\em different} strings $a$
and $a'$.

Let $\ket{\Psi}$ be the entangled state shared by Alice and Bob and define a corresponding density matrix
$\rho=\ket{\Psi}\bra{\Psi}$. Let $\ensuremath{\mathcal{\tilde W}_{\q{r}}}=\{\tilde{W}_{\q{r}}^{\an{r}}\}$ and
$\ensuremath{\mathcal{ W}_{\q{r}}}=\{W_{\q{k}}^{\anp{k}}\}$ be the measurements that they perform when asked
questions $\q{r}$ resp. $\q{k}$ giving answers $\an{r}$ resp. $\an{k}'$. As in Sec. \ref{sec:MIP} we can
assume that these measurements are projective.

The provers pass the consistency test with probability $\pi_2 = \frac{1}{r}\sum_{k=1}^r \pi_2(k)$, where
$$\pi_2(k)= E_{\q{r}}\left[\sum_{\an{r}} \Tr\left(\wtr\otimes\wkk \rho\right)\right]$$
is the probability that the two provers give consistent answers when the verifier has picked $k$ as the
separation point. Conditioned on the fact that they gave consistent answers, they succeed in the classical
test with probability $\pi_1 = \frac{1}{r}\sum_{k=1}^r \pi_1(k)$ where
$$\pi_1(k) =E_{\q{r}}\left[ \sum_{\an{r}} p_q(\an{r}|\q{r},k) V(\an{r}|\q{r})\right]$$
and $p_q(\an{r}|\q{r},k)= \Tr\left(\wtr\otimes\wkk \rho \right)$ is the probability that Alice answers
$\an{r}$ and Bob answers consistently, given that the verifier picked index $k$.

\paragraph{Rounding to a classical strategy:}
Given a strategy for the constructed entangled-prover game $G_c$, we define a strategy for the classical
prover of the original game $G$ in the following way. In round $k$, given the questions to the prover so far
are $\q{k}$ and the prover gave answers $\an{k-1}$, he answers $a_k$ to question $q_k$ with probability
$$p_{class}(a_k|\q{k},\an{k-1})=\frac{\Tr\left(\Id \otimes \wkk \wk{k-1} \cdots \wk{1}\rho\right)}
{\Tr\left(\Id \otimes \wk{k-1} \cdots \wk{1} \rho\right)}$$ (recall that all $\an{k},\an{k-1},\ldots ,\an{1}$
refer to substrings of the same string).
 Note that
$\sum_{a_{k}}p_{class}(a_k|\q{k},\an{k-1})$ could be less than $1$ (we will see from its operational
definition that it is always bounded by $1$). To complete it to a probability distribution we add a special
symbol ``abort" that the prover can send in any round making him lose the game.\footnote{Technically speaking
the extra symbol makes it a different game. We could also have the prover send a random answer whenever
sampling from the complement of the distribution. This can at most increase the prover's winning probability,
so both games have winning probability bounded by $\omega$.}

This probability distribution has the following interpretation. For any operator $A$, denote $A(\rho)=A\rho
A^\dagger$. In the first round the prover in the classical game receives a question $q_1$, and applies the
measurement $\mathcal{{W}}_{\q{1}}$ on Bob's part of $\rho$, answering $a_1$ with probability
$\Tr\left(\Id\otimes \wk{1} \rho\right)=p_{class}(\an{1}|\q{1})$. He is then left with the state
$\frac{\Id\otimes \wk{1} (\rho) }{\Tr\left(\Id\otimes \wk{1} \rho \right)}$. Upon receiving a question $q_2$
in the second round, he measures this state with $\mathcal{W}_{\q{2}}$, answering $a_2$ with probability
$\frac{\Tr\left(\Id\otimes \wk{2} \wk{1} \rho\right)}{\Tr\left(\Id\otimes \wk{1}
\rho\right)}=p_{class}(a_2|\q{2},\an{1})$ if as a result of his measurement he obtains a sequence
$\an{2}=a_1a_2$ consistent with the $a_1$ he had measured in the first round, and an abort symbol in case the
sequence he measures has an $a'_1\neq a_1$. The resulting state in case of non-abortion is $\frac{\Id\otimes
\wk{2}\wk{1} (\rho) }{p_{class}(a_2|\q{2},\an{1})\Tr\left(\Id\otimes \wk{1} \rho \right)}=\frac{\Id\otimes
\wk{2}\wk{1} (\rho) }{\Tr\left(\Id\otimes \wk{2}\wk{1} \rho \right)}$. The prover proceeds similarly at the
subsequent rounds. In other words the prover sequentially performs all the measurements
$\mathcal{W}_{\q{k}}$, and answers according to the resulting distribution, aborting in case the answers he
measures in round $k$ contradict the answers that he has already given in previous rounds.

What is the probability that a fixed sequence of answers $\an{r}$ is given by the prover? We have that
$p_{class}(\an{r}|\q{r})=p_{class}(a_r|\q{r},\an{r-1})\cdot \cdots \cdot p_{class}(a_2|\q{2},a_1)\cdot
p_{class}(a_1|\q{1})$. Because of cancellation, we obtain
$$p_{class}(\an{r}|\q{r})=\Tr\left(\Id\otimes \wk{r} \cdots \wk{1} \rho \right).$$

We will show that this classical strategy is a good one by relating $p_{class}(\an{r}|\q{r})$ to
$p_q(\an{r}|\q{r},r)$ as per the following lemma.

\begin{lemma}\label{lemma:4}
The (weighted) statistical distance between $p_{class}$ and $p_q$ is
$$\Delta(p_{class},p_q)=E_{\q{r}}\left[ \sum_{\an{r}} \Big|p_{class}(\an{r}|\q{r})-p_q(\an{r}|\q{r},r)\Big|\right] \leq 7\,r\,\sqrt{\eps}.$$
\end{lemma}

This lemma is the analogue of Lemmas \ref{lemma:2} and \ref{lemma:3}, and its proof is very similar. Before
proceeding to its proof, we first show how it implies Lemma \ref{lemma:last}. For the total acceptance
probability of the entangled provers we have $1-\eps \leq 1/r\sum_{k=1}^r \min(\pi_1(k),\pi_2(k))$ because
for any index $k$ that is picked by the verifier, we require the provers to succeed in both the Classical
Test and the Consistency Test. This implies that $\pi_1(r)\geq 1-r\eps$, so Bob's  answers can be used to
give correct answers to the Classical Test with probability at least $1-r\eps$, and by Lemma \ref{lemma:4}
this implies that the Classical Test has success probability at least $1-r\eps-7r\sqrt{\eps}$. For
$\eps=\frac{s_2}{r^2}$ for a sufficiently small constant $s_2$ this is more than $s_{IP}$, which implies
Lemma~\ref{lemma:last}.
\end{proof}

\begin{proof}[Proof of Lemma \ref{lemma:4}]
As in the case of three-prover classical entangled games, the fact that Alice's and Bob's answers must be
consistent means that Alice's answers can be used to predict Bob's, so Bob cannot use his share of the
entanglement too much if they are to succeed in the Consistency Test. This means that the action of Bob's
operators $\mathcal{W}$ on the entangled state $\rho$ is close to the identity, at least when the first
prover applies the corresponding $\tilde{\mathcal{W}}$ on his share of $\rho$. The following Claim makes this
explicit and will be used to relate the classical and quantum strategies.

\begin{claim}\label{claim:ipw}
Let the projector $\vt{k}=\sum_{a_{k+1},\ldots,a_r} {\tilde W}_{\q{r}}^{\an{r}}$. The following hold for
every $k\in\{1,\ldots,r\}$:
\begin{align}
&E_{\q{r}}\left[\sum_{\an{k}}\left\|\Id\otimes \wkk (\rho)  - \vt{k}\otimes \wkk (\rho)\right\|_1\right]
\leq 3\sqrt{ 1-\pi_2(k)},\label{eq:pi21}\\
&E_{\q{r}}\left[\sum_{\an{k}}\left\|\vt{k}\otimes \Id (\rho) - \vt{k}\otimes \wkk(\rho)\right\|_1\right]
\leq 3\sqrt{ 1-\pi_2(k)},\label{eq:pi24}\\
&E_{\q{r}}\left[\sum_{\an{k}} \left\| \vt{k-1}\otimes\wkk(\rho)-\vt{k}\otimes\wkk(\rho)\right\|_1 \right]
\leq 1-\pi_2(k).\label{eq:pi23}
\end{align}
\end{claim}

\begin{proof}
Eqs. (\ref{eq:pi21}) and (\ref{eq:pi24}) are a direct application of Lemma \ref{lem:winter}, combined with
the definition of $\pi_2(k)$.
To prove Eq. (\ref{eq:pi23}), note that since $\vt{k-1}\otimes\wkk(\rho)\geq\vt{k}\otimes\wkk(\rho)$, we have
that
\begin{align*}
\left\| \vt{k-1}\otimes\wkk(\rho)-\vt{k}\otimes\wkk(\rho)\right\|_1 &=
\Tr(\vt{k-1}\otimes\wkk(\rho))-\Tr (\vt{k}\otimes\wkk(\rho))\\
& = \sum_{a'_k\neq a_k,a'_{k+1},\ldots,a'_r} \Tr (\tilde{W}_{\q{r}}^{\an{k-1}a'_k \ldots a'_r}\otimes
W_{\q{k}}^{\an{k}} \rho).
\end{align*}
Since $\sum_{\an{r},\anp{k}} \Tr (\tilde{W}_{\q{r}}^{\an{r}}\otimes W_{\q{k}}^{\anp{k}}\rho)=1$,
\begin{align*}
1-\pi_2(k) &= E_{\q{r}}\left[\sum_{\an{r},\anp{k}\neq\an{k}}  \Tr(\tilde{W}_{\q{r}}^{\an{r}}\otimes
W_{\q{k}}^{\anp{k}}\rho)\right]  \geq E_{\q{r}}\left[\sum_{\an{r},a'_k\neq a_k} \Tr
(\tilde{W}_{\q{r}}^{\an{r}}\otimes W_{\q{k}}^{\an{k-1},a'_k}\rho)\right]
\end{align*}
which concludes the proof.
\end{proof}
Observe that for any set of orthogonal projectors $\{W^a\}$ we have that $\sum_{a} \| W^a \sigma_1 W^a - W^a
 \sigma_2 W^a \|_1 \leq \|\sigma_1-\sigma_2\|_1$ for any two matrices $\sigma_1$, $\sigma_2$. Using this
 successively for the sets $\{W_{\q{2}}^{\an{2}}\}_{a_2}$, ..., $\{W_{\q{r}}^{\an{r}}\}_{a_r}$, from Eq.
 (\ref{eq:pi21}) with $k=1$ we get
\begin{align*}
E_{\q{r}}\left[ \sum_{\an{r}} \left\| \Id\otimes \wk{r}\cdots\wk{1}(\rho) - \vt{1}\otimes \wk{r}\cdots
\wk{1}(\rho)\right\|_1 \right] &\leq 3\sqrt{1-\pi_2(1)}.
\end{align*}
Similarly, from Eq. (\ref{eq:pi24}),
\begin{align*}
E_{\q{r}}\left[ \sum_{\an{r}} \left\| \vt{1}\otimes \wk{r}\cdots \wk{1}(\rho)-\vt{1}\otimes \wk{r}\cdots
\wk{2}(\rho)\right\|_1 \right] &\leq 3\sqrt{1-\pi_2(1)}
\end{align*}
and from Eq. (\ref{eq:pi23}) with $k=2$
\begin{align*}
E_{\q{r}}\left[ \sum_{\an{r}} \left\| \vt{1}\otimes \wk{r}\cdots \wk{2}(\rho)-\vt{2}\otimes \wk{r}\cdots
\wk{2}(\rho)\right\|_1 \right] &\leq 1-\pi_2(2)
\end{align*}
Repeating these operations for each $k$, adding the equations and using triangle inequality finally yields
\begin{align*}
E_{\q{r}}\left[ \sum_{\an{r}} \left\|\Id\otimes \wk{r}\cdots\wk{1}(\rho) - \vt{r}\otimes \wk{r}(\rho)\right\|_1 \right]&\leq 6\sum_{k=1}^r \sqrt{1-\pi_2(k)} + \sum_{k=2}^r (1-\pi_2(k)) \\
&\leq 7 r\sqrt{1-\pi_2}
\end{align*}
using concavity of the function $\sqrt{1-x}$. Since $\vt{r}={\tilde W}_{\q{r}}^{\an{r}}$, the lemma follows
because the trace distance is an upper bound on the variation distance of the probability distribution
resulting from making any measurement on these two states.
\end{proof}

\section{Conclusions and Open Questions}\label{sec:rest}

We have established that it is \NP-hard to approximate the value of
both two-prover quantum entangled games and three-prover classical
entangled games. These results leave open the case of {\em two}-prover
one-round {\em classical} entangled games. Can our techniques be
extended to this case?

The other obvious question is whether we can improve the
inapproximability ratio to better than an inverse polynomial in the
number of questions. Are there additional tests that further limit the
advantage provers can obtain by sharing entanglement? For example, in
the case of classical entangled games, does it help to add more than
three provers? In particular, if there are as many provers as there
are questions, then sharing entanglement does not help, even if the
verifier only talks to two provers chosen at random.

In very recent work \cite{KKMV07} a
subset of the authors obtain parallelization results for the case of
quantum multi-round entangled games, showing that any such game with
$k$ provers and $r$ rounds can be parallelized to a $3$-turn game with
$k$ provers at the expense of a $\poly(r)$ factor in the value of the
game. Moreover, such a game can be parallelized to $2$ messages, or
$1$ round, by adding a $(k+1)$-st prover. We do not know whether it is
possible to parallelize quantum entangled games from three to two
messages without adding an additional prover.

There are a number of other important questions that our work does not
address.  Can we prove \emph{upper} bounds on the hardness of
computing the value of entangled games? It is instructive here to
compare to the case where the provers share no-signalling
correlations, where there is an efficient linear-programming algorithm
to compute the value of a game~\cite{preda:_nosig}.\footnote{The reason that
our proof does not work for no-signalling provers is that there is no
notion of a partial measurement of a no-signalling probability
distribution, so the classical strategy we use in our proofs cannot be
defined.} In the quantum case, it is still not known whether the
decision problem corresponding to finding the value of an
entangled-prover game is recursive! The issue is that we are not
currently able to prove any bounds on the amount of entanglement
required to play a game optimally, even approximately.

\section{Acknowledgments}
We thank  Tsuyoshi Ito, Jaikumar Radhakrishnan, Oded Regev, Amnon Ta-Shma, Mario Szegedy  and Andy Yao for
helpful discussions and John Watrous for pointing out that the optimal quantum value of a game might not be
achieved with finite dimensional entanglement.


\newcommand{\etalchar}[1]{$^{#1}$}

\end{document}